\documentclass[aps,pre,amssymb,amsmath,preprint]{revtex4-1}
\usepackage{natbib}
\usepackage{graphicx,color}

\begin{document}
\title{Relation between boundary slip mechanisms and water-like fluid behavior}%

\author{Patricia Ternes}%
\email{patricia.terdal@gmail.com}%
\affiliation{Instituto de F\'isica, Universidade Federal do Rio Grande do Sul,
Caixa Postal 15051, 91501-970, Porto Alegre, RS, Brazil}%

\author{Evy Salcedo}%
\affiliation{Coordenadoria Especial de F\'isica, Qu\'imica e Matem\'atica, 
Universidade Federal de Santa Catarina, Rua Pedro Jo\~ao Pereira, 150, 88905-120, 
Ararangu\'a, SC,  Brazil}%

\author{Marcia C. Barbosa}%
\affiliation{Instituto de F\'isica, Universidade Federal do Rio Grande do Sul,
Caixa Postal 15051, 91501-970, Porto Alegre, RS, Brazil}%

\date{\today}%

\begin{abstract}
The slip of a fluid layer in contact with
a  solid confining surface is investigated
for different temperatures and densities 
using molecular dynamic simulations.
We show that for an anomalous water-like
fluid the slip goes as follows:  for low
levels of shear, the   defect slip
appears  and is related with the particle 
exchange between the fluid layers; at high
levels of shear, the global slip occurs and 
is related to the  homogeneous distribution 
of the fluid in the confining surfaces. 
The oscillations in the transition velocity
from the defect to the global slip is shown
to be associated with changes in the 
layering distribution in the anomalous 
fluid.
\end{abstract}
\maketitle
\section{Introduction}
The no-slip condition is the assumption that 
the fluid velocity is zero when in contact 
with the solid confined geometry. For a 
macroscopic flow this is a trustworthy
boundary condition, and it is fundamental 
for the continuum theory validity. For 
confined geometries the hydrodynamic 
equations are no longer valid and, in this 
case, the use of the no-slip boundary 
condition is at least questionable. Many 
experimental~\cite{WAT99,BAU01,ZHU01,ZHU02,COT05,NET05,JOS05,LI14},  
theoretical and computational 
results~\cite{BOC94,VIN95,SOK01,BOC07,NIA10,BHA15,WAG17} 
report that there are several flow 
boundary conditions  consistent with the 
fluid behavior and 
mobility~\cite{LIC04,LIC07,MAR08I,MAR08II}
beyond the no-slip boundary condition.
The amount  of slip is usually measured  through
the magnitude of the slip length, defined as 
the ratio between the shear rate
and  the slip velocity~\cite{LIC04,LIC07,MAR08I,MAR08II}. 
For most liquids the slip length increases 
with the shear rate and stabilizes at 
$v_0$~\cite{MAR08II}. Its value, however, 
depends on the thickness of the confining 
system in a non trivial way. For apolar 
materials, such as the hexane~\cite{CHE02} 
and the n-decane~\cite{LIC07,MAR08II}, 
the slip length increases with the film 
thickness. However, for the 
polyamide-6,6~\cite{ESL10} and for 
water~\cite{THO08} the slip length 
decreases with the increase of the film.
Complementary the behavior of the shear  
viscosity gives the slip length.
For the n-decane the viscosity increases
with the increase of the film thickness~\cite{MAR06}
while for polyamide-6,6~\cite{ESL10} 
the viscosity decreases with the increase 
of the film.

For water the situation is even more complex.
Confined water in microchannels presents a slip
length in the order of 
nanometers~\cite{TRE02,BON02,CHO03,VIN09,XUE14}
and therefore no-slip boundary conditions is
no applicable.  As the channel size decreases, 
water mobility increases~\cite{MAJ05,HOL06,QIN11,TER17}. 
The slip length for nanochannels becomes 
of the order of micrometers what implies
that the use of no-slip boundary conditions 
could be problematic. 

Even thought the qualitative behavior of 
the slip length is known, the specific
value of the slip length is widely 
scattered. In the particular case of 
water, it depends on the surface energy 
and roughness, the fluid temperature and
density~\cite{KAN12,KOP89,THO90,THO97,BAR99,PI00,CIE01,CHO03,JOL06,PRI06,PRI07}.
Then, a strategy to understand this qualitative behavior
is to explore the mechanism behind the 
change of the slip with the shear rate,  
if the slip occurs through one single 
process or if it involves a number of 
steps which depend on the velocity. 
In the case of apolar Lennard-Jones  
fluids the slip  changes through  
two mechanisms: the  defect slip 
and the global 
slip~\cite{LIC04,LIC07,MAR08I,MAR08II}. 
The transition from the defect to the 
global slip occurs at a shear rate $v_0$. 
Therefore for a given fluid and wall, the
slip length depends on the temperature 
and density  from the behavior of $v_0$. 
Unfortunately very little is known about 
the behavior of $v_0$.

Here we add another component to this 
already complex problem. We explore 
the mechanism behind the liquid slip
 in the case of anomalous water-like 
fluids. An anomalous fluid is characterized
by having a maximum in the density versus
temperature at fixed pressure and a maximum
and a minimum in the diffusion coefficient 
versus pressure at constant 
temperature~\cite{NET01,XU05,XU06,OLI06I,OLI06II,BAR09}. 
Under high confinement, these fluids exhibit 
additional anomalous behaviors and new 
phases~\cite{BOR14II,BOR14I}.
When an anomalous fluid is nanoconfined
the thermodynamic and  dynamic 
properties differ from the properties 
observed in the bulk~\cite{MAJ05,HOL06,QIN11,GAV17}.  
For instance, for the bulk system
the fluid is described as homogeneously 
distributed. This is not the case for 
the confined systems. The water-like fluid
forms layers which depend on the 
film thickness~\cite{BOR14II,BOR14I,NEE16,GAL10,FAR16,DEM17}.
Due to the layering, particles have 
different behavior at different layers,  
what allow for the anomalous flux.
observed in confined water-like materials.

In this paper we  investigate the slip mechanism
of a water-like fluid. After testing
the fluid for  the defect slip and global slip
transition at $v_0$, we study the connection 
between the behavior of $v_0$ at different 
temperatures and densities with the structure 
and dynamics of the layers. 

The water-like fluids is
modeled by an  effective potential with two length
scales separated by an energy barrier. The use of
an effective potential allows us to explore a large
range of the temperature versus density phase diagram.
Molecular dynamics simulations of the planar Couette flow
for this anomalous fluid test the presence
of defect and global slip at different densities and temperatures.
This paper goes as follows:  section II  presents the model and
simulation details,   section III shows the results and
the section IV  has the conclusions.

\section{The model, simulation and methods}
\subsection{The model}

\begin{figure}[h!]
  \begin{center}
  \includegraphics[width=6.250in ]{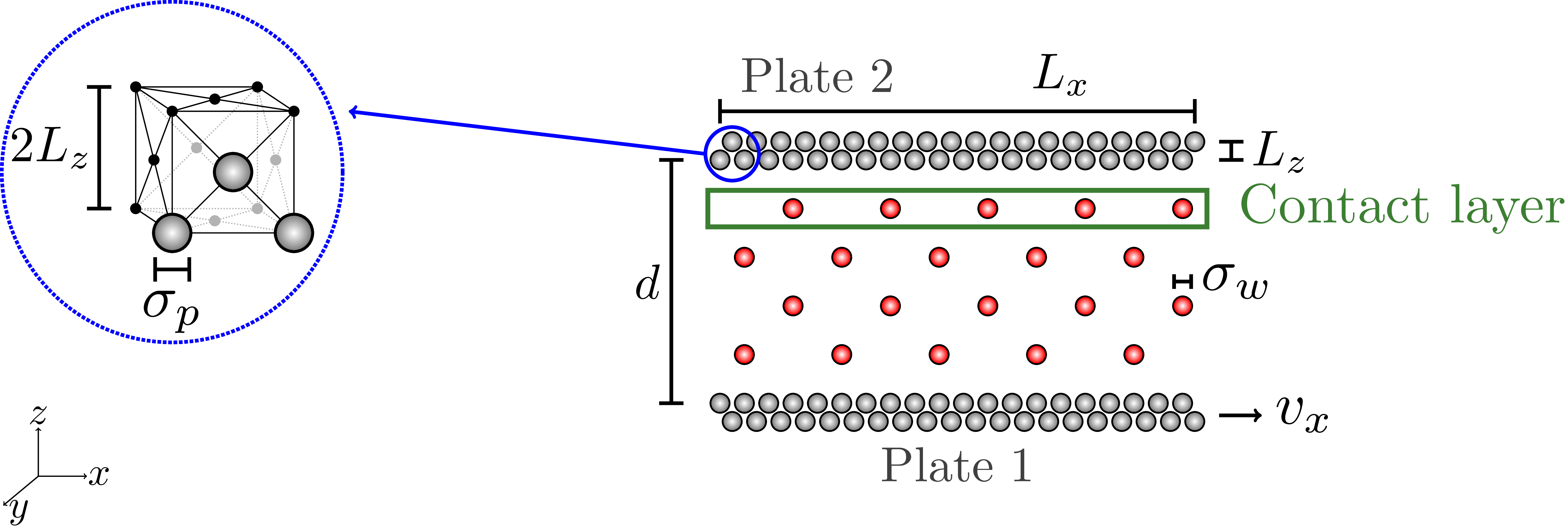}
  \caption{(Color online) Schematic representation of 
water-like fluid confined within parallel plates.
           The y-direction is omitted.}
  \label{ref_fig1}
  \end{center}
\end{figure}

The water-like fluid is  confined in a 
planar Couette geometry shown in the 
figure~\ref{ref_fig1}. Each plate is 
formed by $N_p=676$ spherical particles
of  diameter $\sigma_p$ organized in 
two planar layers, forming a face 
centered cubic lattice, with 
$L_x=L_y=20.2\sigma_p$ and $L_z=0.7\sigma_p$, 
as shown in the figure~\ref{ref_fig1}. 
The separation between the plates, or
channel height, is $d$. The liquid is 
sheared by moving the bottom bounding 
wall (plate 1 in the figure~\ref{ref_fig1}) 
with the  speed $v_x$ while the 
top bounding wall (plate 2 in the 
figure~\ref{ref_fig1}) is held fixed. The 
contact layer, is composed by fluid 
particles whose centers of mass 
lie between the plate 2 and the 
first minimum in the density profile.

\begin{figure}[h!]
  \begin{center}
    \includegraphics[height=2.9in ]{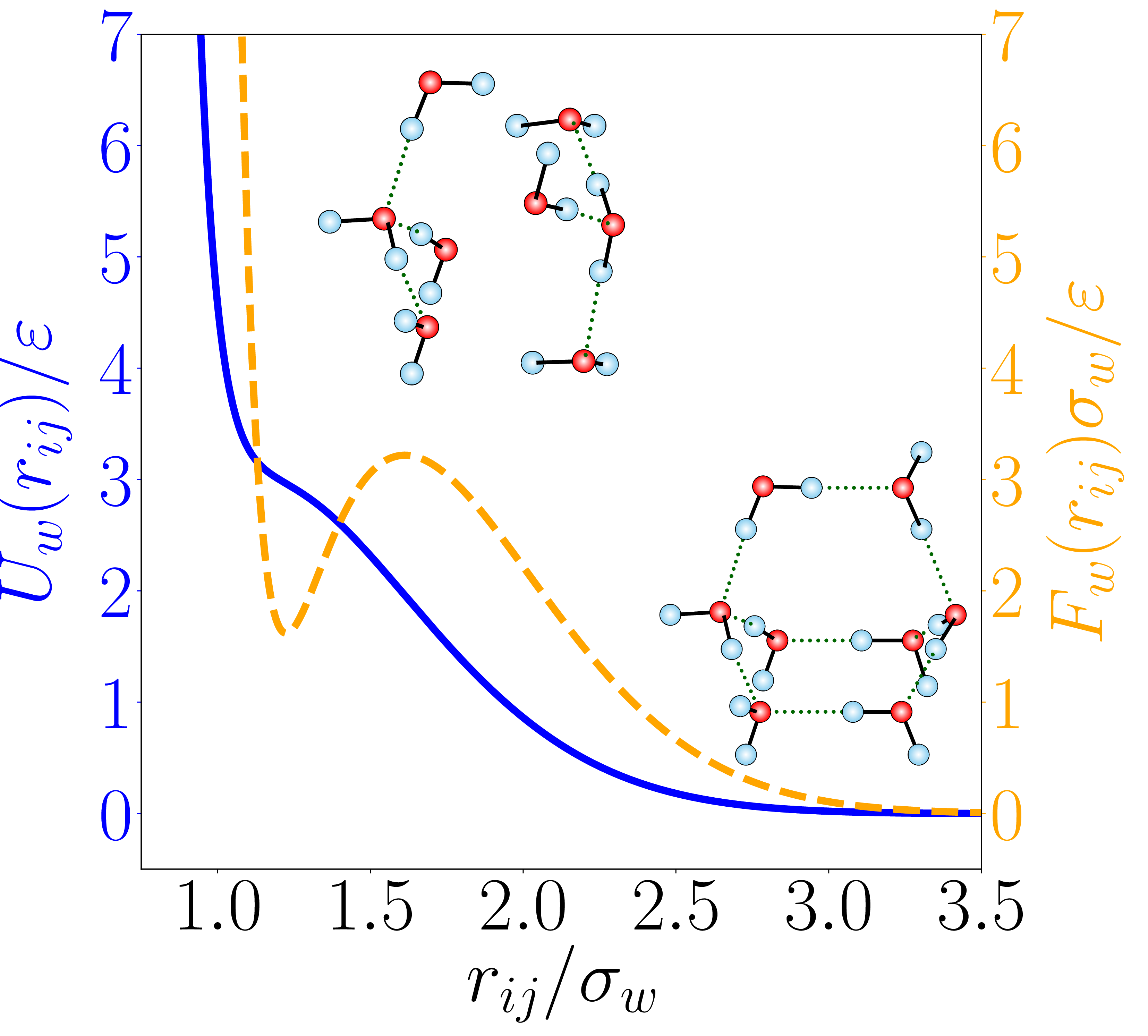}
  \end{center}
  \caption{(Color online) Left axis: isotropic effective potential  
            as a function of the particle separation (blue solid line).
            Right axis: force related to the effective potential as a 
            function of the particle separation (orange dashed line).}
  \label{ref_fig2}
\end{figure}

The particles at the plates are tethered
to its lattice site by a linear spring
with constant 
$k^* = k\left[\sigma_w^2/(m\varepsilon)\right]^{1/2}=50$
and characteristic excursion 
$\xi=\sqrt{k_BT/k}$~\cite{SEG13}, where $k_B$ is 
the Boltzmann constant, and $T$ is the plate 
temperature. The particles at the plates also
interact with each other via a standard
Lennard-Jones (LJ) 12-6 potential with 
$\varepsilon$ depth and $\sigma_p$~\cite{JON24,JON31}. 
The fluid is modeled by $N_w=500$ identical 
water-like particles with diameter 
$\sigma_w=\sigma_p$. The fluid particles 
interact through a core-softened potential 
given by~\cite{OLI06II,OLI06I}
\begin{equation}
   \dfrac{U_w\left( r_{ij}\right)}{\varepsilon}  
   = 4 \left[ \left(\frac{\sigma_w}{r_{ij}}
   \right)^{12} - \left(\frac{\sigma_w}{r_{ij}}
   \right)^{6}\right] + u_0 \exp \left[-\frac{1}
   {c_0^{2}}\left(\frac{r_{ij} - r_{0}}{\sigma_w} 
   \right)^{2} \right]\; . 
\end{equation}
This potential presents two length scales which
consists of a standard Lennard-Jones 12-6 
potential (LJ) plus a Gaussian centered at 
$r_0$,  with width $c_0$ and depth $u_0$, where
$r_{ij} = |\vec{r}_i-\vec{r}_j|$ is the distance 
between fluid particles $i$ and 
$j$~\cite{OLI06II,OLI06I}. Varying the parameters 
$u_0$, $c_0$, $r_0$ and $\sigma_w$ this potential 
can represent a whole family of intermolecular 
interactions. In this work the chosen parameters 
are $u_0 = 5.0$, $c_0=1.0$ and $r_0=0.7\sigma_w$. 
For these parameters the potential presents one 
scale at $ r_{ij} \approx 1.2\sigma_w$ and other 
scale at $ r_{ij} \approx 2\sigma_w$, being each 
scale related to the interaction between two 
water tetramer clusters, as shown in the 
illustration of the force in the  
figure~\ref{ref_fig2}~\cite{GOR93,STI93}.
The bulk system of spherical particles 
interacting through this potential
exhibits diffusion, structural and
density anomalous behavior observed 
also in bulk water~\cite{OLI06II,OLI06I}.

The particles of this water-like fluid 
interact with the wall particles through 
the purely repulsive potential given by 
the Weeks-Chandler-Andersen Lennard-Jones 
(WCA) potential~\cite{WEE71,FRE96}
\begin{equation}
  U_{wp}(r_{ij}) = 
  \left \{ \begin{matrix} U_{LJ}(r_{ij}) - 
  U_{LJ}(r_c) &\mbox{;} \qquad r_{ij} \le r_c  \\ 
  0 & \mbox{;} \qquad  r_{ij}>r_c\end{matrix}\right. \;,
  \label{Pot_WCA}
\end{equation}
where $U_{LJ}$ is a standard 12-6 LJ and $r_c$ 
is cutoff distance ($r_c = 2^{1/6}\sigma_{wp}$). 
The effect radius, $\sigma_{wp}$, is determined 
through Lorentz-Berthelot mixing rule 
($\sigma_{wp}=(\sigma_{p}+\sigma_{w})/2$) and is 
used when one fluid particle is interacting
with one wall particle~\cite{ALL87}. The 
repulsive fluid-plate interaction causes an 
excluded volume, therefore the fluid effective 
density will be $\rho=N_w/[(d -\sigma_{wp})
L_xL_y]$~\cite{KUM05,KUM07}.

\subsection{The simulations}
The system was studied by Molecular Dynamics
(MD) simulations at constant $NVT$ 
through a homemade program. Nos\'e-Hoover 
heat-bath with coupling parameter $Q=2$ 
was applied at the plates particles in order
to maintain the temperature fixed~\cite{HOO85,HOO86}.
The system was analyzed for different 
densities and heat-bath temperatures. 
The temperature varies from $T^*=k_BT/\varepsilon=0.025$
up to $T^*=0.650$, and  $\xi$, varies from $0.022$ 
up to $0.110$. $N$ is fixed and  the density decreases 
by increasing the distance between the plates from  
$d^*=d/\sigma_w=3.8$ to $d^*=9.8$.
The initial configuration of the confined 
fluid is set on the solid state and, without 
shear ($\vec{v}_x=0$), further equilibrated 
over $5\times 10^5$ steps. Then, in order to
obtain the temperature versus density phase 
diagram of the confined system, $2\times 10^6$ 
steps were performed. The transversal pressure, 
$P$, is computed analogously to bulk pressure~\cite{ZAN00}
\begin{equation}
 P = \rho k_B T + \dfrac{1}{V}\left\langle \nu_{\perp} \right\rangle\; , 
\end{equation}
where $\nu_{\perp}$ is the transversal Virial expression,
\begin{equation}
  \nu_{\perp} = -\sum_{i}\sum_{j>i}\dfrac{z_{ij}^{2}}{r_{ij}}
  \left(\dfrac{\partial U_w\left(r_{ij}\right)}{\partial r_{ij}}\right)\; . 
\end{equation}

Next, the bottom bounding wall moves with a 
constant speed $\vec{v}_x$. For each density 
and temperature, several simulations 
with wall velocities varying from low shear 
levels, $v_x^*=v_x \left(m/\varepsilon\right)^{1/2}= 0.001$,
up to high shear levels, $v_x^*=15.0$ (where 
the bottom wall velocity is about five times  
greater than the fluid thermal velocity) were 
carried out. The fluid heats up due to shear 
and the system reaches a new equilibrium 
temperature after $3\times10^5$ steps. 
Since the equilibrium temperature of the 
fluid depends on shear level, the temperature 
used in the graphs is the heat-bath obtained 
from the  thermostat fixed at the wall.
After the equilibration, additional 
$8\times 10^6$ steps were performed to store
physical quantities for the system with shear. 
The structure of the water-like fluid in the
contact layer was analyzed through the parallel
radial distribution function, $g_{\|}(r_{xy})$. 
This distribution function is defined as~\cite{KUM05}
\begin{equation}
  g_{\|}(r_{xy}) \equiv  \dfrac{1}{\rho^2V}\sum_{i \neq j} 
\delta(r_{xy} - r_{ij})
  [\theta (|z_i - z_j|)  - \theta (|z_i - z_j| - \delta z)]\; ,
\end{equation}
where $r_{xy}$ is the parallel distance between
particles, and $\theta(z)$ is the Heaviside 
function which limits the particle sum in a layer 
of thickness $\delta z$.  The fluid 
structure was also analyzed through the translational 
order parameter, defined as~\cite{ERR01}
\begin{equation}
  t \equiv \int_{0}^{\zeta_c} | g_{\|}(\zeta) -1| d\zeta\;,
\end{equation}
where $\zeta_c=0.5L_x\rho_{l}^{1/2}$ is the cutoff 
distance set to half of the simulation box times
density of the contact layer, and $\zeta=r_{xy}(\rho_{l})^{1/2}$ 
is the distance $r_{xy}$ in units of the mean 
interparticle separation in the parallel direction.
The translational order parameter measure how 
structured is the system. For ideal gas, $t=0$ and 
for more structured phases, $t$ increases.

The equations of motion were integrate with a 
time step $\delta t^* = \delta t\left[\varepsilon/
(m\sigma_w^2)\right]^{1/2}= 0.0025$, and five 
independent runs were used to evaluate the confined 
anomalous fluid properties. All the quantities are 
given in Lennard-Jones units~\cite{ALL87}
and, for simplicity, the symbol (*) employed in the 
dimensionless quantities is excluded. 

\subsection{The slip boundary conditions}
Usually confined systems are analyzed employing 
no-slip boundary condition in which the mean 
velocity of the fluid particles in the contact 
layer is zero. Even though the no-slip boundary 
condition is  good to describe confinement up
to microchannels~\cite{TRE02,BON02,CHO03,VIN09,XUE14}, 
this might not the case for nanoconfined 
geometries~\cite{MAJ05,HOL06,QIN11,TER17}. 
Different slip conditions mechanisms might 
occur as the relative velocity between the 
fluid and the wall is changed. For the planar
Couette flow two boundary slip mechanisms are 
predicted for non-anomalous fluid: the defect 
slip and the global slip~\cite{LIC04,LIC07,MAR08I,MAR08II}. 
The defect slip depends on the  local and 
ordered hops of the fluid particles at the 
contact layer. These hops occur due to the 
presence of disorder in the ground state of the 
wall-fluid interaction which obeys an Arrhenius 
dynamics. The global slip occurs when all fluid 
particles of the contact layer are in movement 
detached from the wall.

It is possible to verify the occurrence of these 
boundary slip mechanisms by analyzing the average 
particle motion. In the no-slip condition the particles 
oscillate around the minimum of the ground state 
of the particle-wall interaction. In this case the 
fluid particles in the contact layer have no 
preferential direction of movement. For the slip 
condition the movement of fluid particles in the 
contact layer is in the driven direction. Then,  to 
compute this move, we compute the probability of 
one particle moving in the driven direction, 
$P_{DD}$, defined as 
\begin{equation}
 P_{DD} = \dfrac{100}{S} \sum_{i=1}^{S}\left[ 
 \dfrac{\sum_{j=1}^{N_{CL}}\left[x_j(i) - 
 x_j(i-1)\right]\sigma_{ij}} {\sum_{j=1}^{N_{CL}}
 |x_j(i) - x_j(i-1)|} \right]\;,
\end{equation}
where $S$ is the number of simulation steps, $N_{CL}$
is the number of fluid particles in the contact 
layer, $x_j(i) - x_j(i-1)$ is the displacement 
of particle $j$ between the steps $i-1$ and $i$, 
and $\sigma_{ij}$ is a piecewise function. If the 
displacement is in the driven direction ($x_j(i) 
- x_j(i-1) \ge 0$), then $\sigma_{ij}=1$, and if the 
displacement is in opposite direction ($x_j(i) - 
x_j(i-1) < 0$), then $\sigma_{ij}=0$. If $P_{DD}$ 
is close to $50\%$ and the contact layer is 
stationary, the no-slip boundary condition are 
valid. As the fluid particles hop from one site to 
another, $P_{DD}>50\%$, the system is in the defect 
slip boundary condition and the particles move 
in one direction. For $P_{DD}$  close to $100\%$ 
the particles at the contact layer move in the 
driven direction and the system is in the global 
slip boundary condition. Then the transition
between the no-slip condition to the defect and 
global slip conditions is identified by the 
transition of the logistic function
\begin{equation}
 P_{DD} = 100 - \dfrac{50}{1 + \left(v_x/v_0 \right)^{\alpha}}\;,
 \label{ref_eq8}
\end{equation}
where $\alpha$ is the steepness of the curve 
that is related with the necessary velocity
increases to promote the transition between 
the defect slip to global slip, and $v_0$ is 
the logistic midpoint that is related with 
the bottom  wall velocity that promote the 
global slip. 

\section{Results and discussion}
\begin{figure}[h!]
  \begin{center}
    \includegraphics[width=4in ]{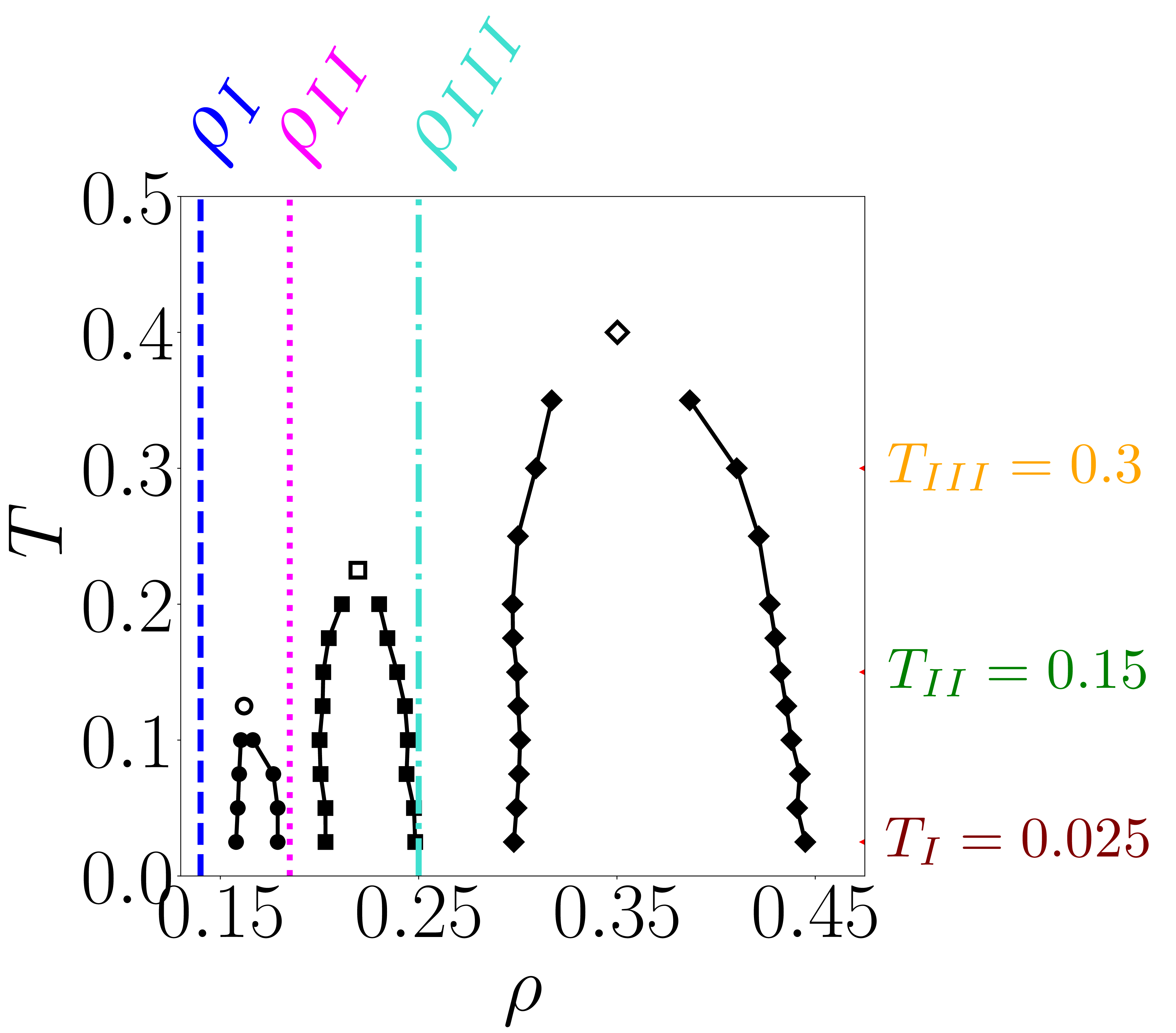}
  \end{center}
  \caption{(Color online) Temperature versus density phase 
diagram without shear. 
          The black solid lines 
	  represents the regions of first order phase transitions 
that ends in three 
	  critical points (empty symbols). The 
temperatures $T_I$, $T_{II}$ and 
	  $T_{III}$, and the densities $\rho_I$, $\rho_{II}$ and 
$\rho_{III}$, 
	  indicate values used in next results.}
  \label{ref_fig3}
\end{figure}

In order to understand the effect of different 
boundary conditions on the behavior of the
water-like fluid, first we obtained the 
behavior of the system at no-slip boundary 
condition (without shear). For 
this system, the pressure versus density 
phase diagram presents isochores monotonic 
with the temperature above $T^*=0.400$. Below
$T^*=0.400$  van der Waals loops indicate 
the presence of a coexistence between two 
phases. The coexistence densities were then 
obtained using the Maxwell construction, and 
the critical points are given by $d^2P/d\rho^2=0$. 
The temperature versus density phase diagram  
in the figure~\ref{ref_fig3} summarizes this 
information.
The regions inside the curves in figure~\ref{ref_fig3}
represent the coexistence between the various 
two-dimensional liquid, liquid crystal and crystal
phases~\cite{BOR14I}. The empty symbols in the 
figure~\ref{ref_fig3} are the critical
points. For zero shear the phase diagrams
obtained using the thermostat at the  wall or at
the fluid are equal~\cite{BOR14I}. 

\begin{figure}[h!]
  \begin{center}
    \includegraphics[width=6.45in ]{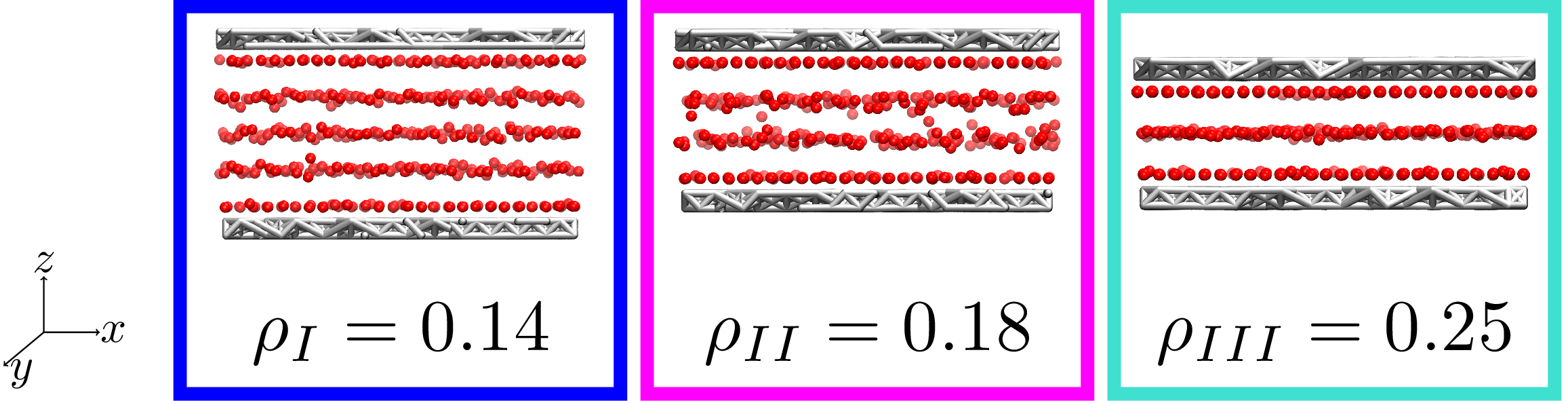}
  \end{center}
  \caption{(Color online) Snapshots of water-like fluid confined 
between parallel plates, 
           without shear, at $T_I=0.025$.
           To simplify viewing, the y-direction is omitted.}
  \label{ref_fig4}
\end{figure}

The confined water-like system is characterized by the
presence of planar layers. The number of layers
dependents on the film thickness, as illustrated in 
figure~\ref{ref_fig4}.
Since in our system the number of particles is
kept fixed, the change of the film thickness
is equal to the change of density. For 
temperatures from $T=0.025$ 
to $0.65$ and for the density
$\rho_I=0.14$ (dashed line in the 
figure~\ref{ref_fig3}), the fluid 
forms five layers; for
the density $\rho_{II}=0.18$ (dotted line in the figure~\ref{ref_fig3}), the
 fluid is
structured in four layers; and for the density $\rho_{III}=0.25$ 
 (dot-dashed line in the
figure~\ref{ref_fig3}), the fluid forms three layers.

\begin{figure}[h!]
  \begin{center}
    \includegraphics[width=3in ]{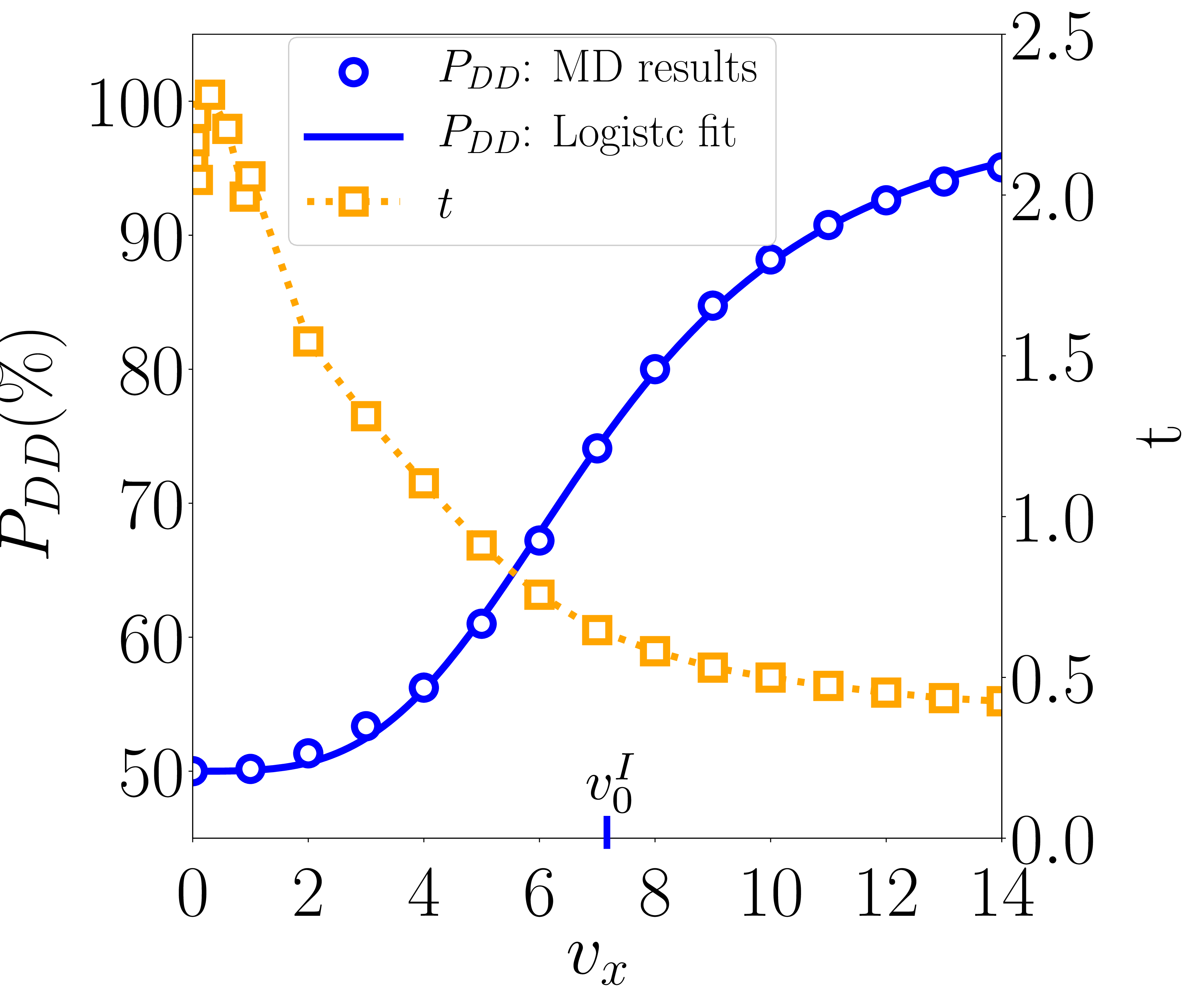}
  \end{center}
  \caption{(Color online) Left axis: Probability of the particles at the 
 contact layer to move
           in the driven direction as a function of the bottom
 wall velocity. 
           The blue circles are results from simulations for
thermal bath temperature, $T_I = 0.025$, and density
of the fluid,
           $\rho_I = 0.14$. The blue solid line is a logistic fit.
           Right axis: translational order parameter as a function of 
           bottom wall velocity for $\rho_I$ at $T_I$. }
  \label{ref_fig5}
\end{figure}

We also explored the effect of different boundary conditions.
In particular, we studied the behavior of the fluid as the bottom wall moves
for systems with 
different fluid densities and several thermal bath temperatures. 
The figure~\ref{ref_fig5} 
illustrates the probability of particle move
in the driven
direction as a function of the bottom wall speed. We identified three 
characteristic regions. The  no-slip condition is valid 
for small velocities of the bottom wall, $50\%$ of 
the particles move in driven direction while the
another $50\%$ move in the opposite 
direction. For the temperatures of $T_I = 0.025$ and $\rho_I = 0.14$,
the  condition of no-slip is valid for velocities up to $v_x<2$.
As the bottom wall velocity increases, $P_{DD}$
increases leading to defect slip condition 
in which a few more  particles move in the
same direction of the bottom wall. As $v_x$ increases even further, the 
system reaches the global slip condition when $v_x>v_0$, which implies that most 
fluid particles in contact layer move in the 
same direction as the bottom wall. 
The transition between the boundary conditions  
is characterized  by  the logistic equation (eq.~\ref{ref_eq8}) illustrated by 
the solid line in the figure~\ref{ref_fig5}.
For $T_I$ and $\rho_I$, the logistic fit is given by
$v_0^I = 7.15(4)$ and $\alpha^I = 3.39(7)$. 
A similar graph is also observed for higher densities and temperatures.
This behavior
was also observed for a Lennard-Jones-like fluid
for one specific temperature and density~\cite{MAR08II}.

\begin{figure}[h!]
  \begin{center}
    \includegraphics[width=2.8in ]{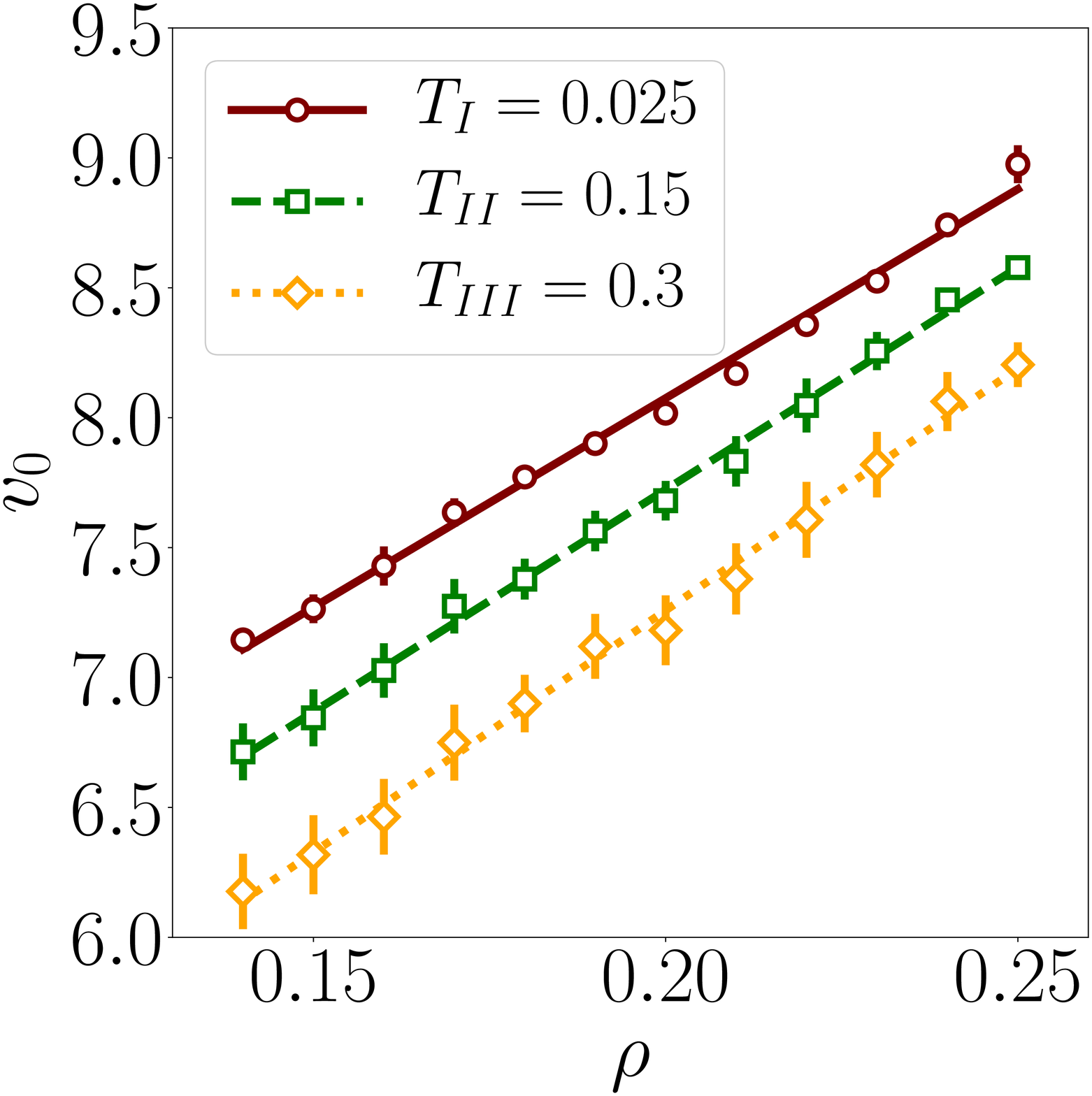}
  \end{center}
  \caption{(Color online) Logistic midpoint as a function of 
density for three different 
           temperatures, where the symbols are results from 
simulations and the lines 
           are linear fits.}
  \label{ref_fig6}
\end{figure}

In order to understand how the slip condition is affected by 
the thermodynamic state of the anomalous fluid,
the behavior of $v_0$ and $\alpha$ are analyzed for different temperatures
and densities (in our system this implies thickness of the film).
The figure~\ref{ref_fig6} shows the behavior of the logistic
 midpoint,  $v_0$, defined
by the equation~\ref{ref_eq8} as a function of the density for 
distinct temperatures 
indicated in the figure~\ref{ref_fig3} ($T_I, T_{II}$ and $T_{III}$). This 
result shows 
that $v_0$ increases with the density at a fixed temperature.

\begin{figure}[h!]
  \begin{center}
    \includegraphics[width=2.8in]{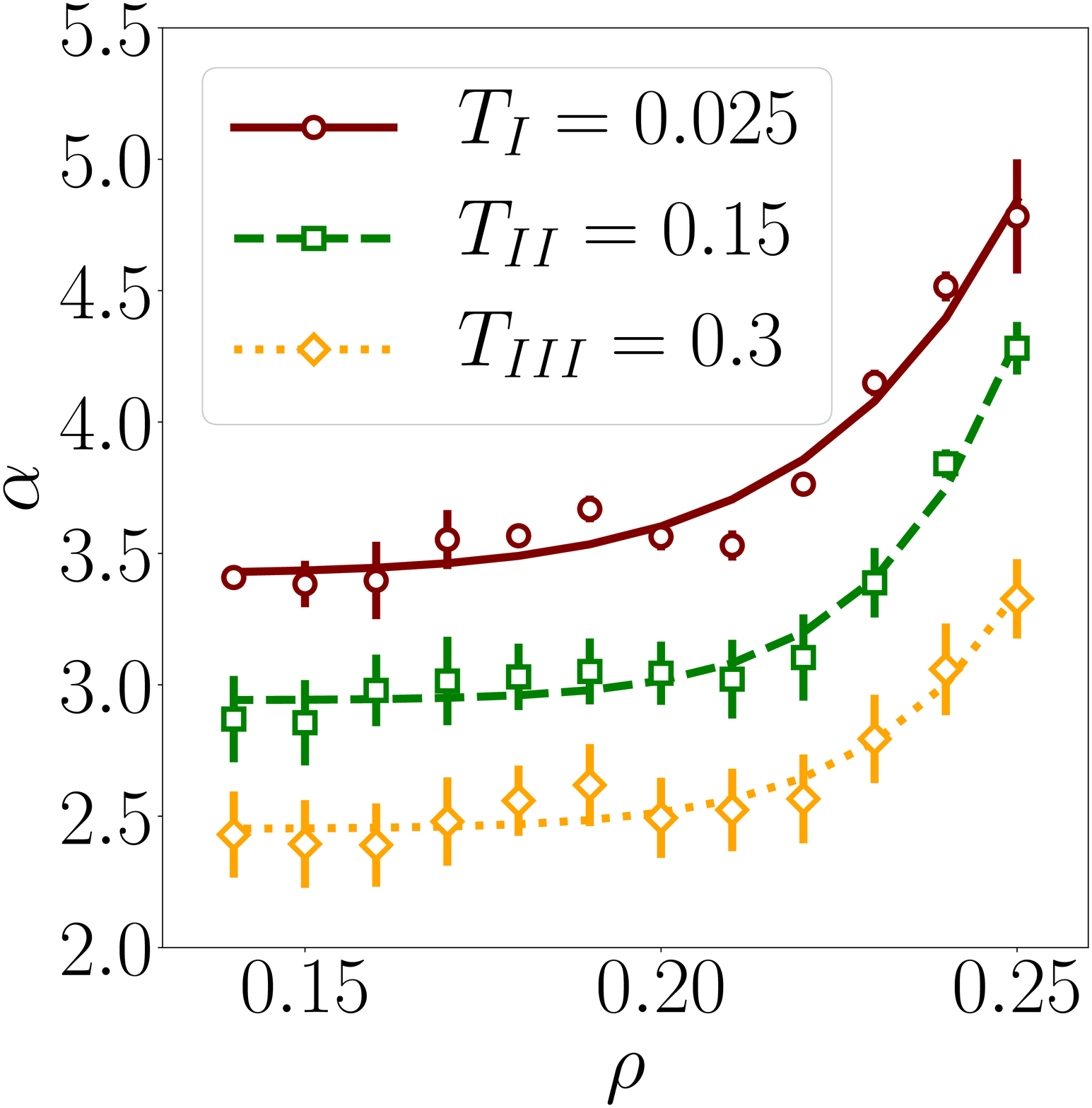}
  \end{center}
  \caption{(Color online) Logistic steepness as a function of 
density for three different temperatures,
           where the symbols are results from simulations and 
the lines are power law 
           curve fittings.}
  \label{ref_fig7}
\end{figure}

At low temperatures the particles
at constant density 
are more structured and higher kinetic energy, larger $v_0$, from
the moving wall
would be needed for the transition to the global slip. Similarly, at constant
temperature, as the system becomes more dense, it is also more 
structured~\cite{BOR14I}, therefore, would require higher value of $v_0$ for the 
transition to the global slip.
The logistic steepness, $\alpha$, defined by equation~\ref{ref_eq8}, versus
$\rho$ for the
temperatures $T=0.025$, $0.15$, $0.3$ is shown in the figure~\ref{ref_fig7}. 
For low densities the value is almost constant, and
for high values of the  densities the $\alpha$ value increases in a power law 
behavior. 
%
%

An
oscillatory behavior
around a line is observed for both $v_0$ and $\alpha$ as a function of 
density.
The oscillatory behavior  presented in the 
figures~\ref{ref_fig6} and~\ref{ref_fig7} 
occurs for the same densities in
both figures.  
The oscillations occur for densities from $0.25$ to $0.20$
and from $0.17$ to $0.16$,
where 
the number of layers shown in the figure~\ref{ref_fig4}
change from three to four, and four to five layers, respectively.
So, even though the qualitative behavior of $P_{DD}$ (fig.~\ref{ref_fig5})
is also observed for Lennard-Jones-like fluids, 
the small oscillations 
in $v_0$ and $\alpha$ versus density indicate that 
the values assumed by $v_0$ and $\alpha$ are related with
the unusual structures assumed by water-like fluid under confinement.

\begin{figure}[h!]
  \begin{center}
    \includegraphics[height=2.7in]{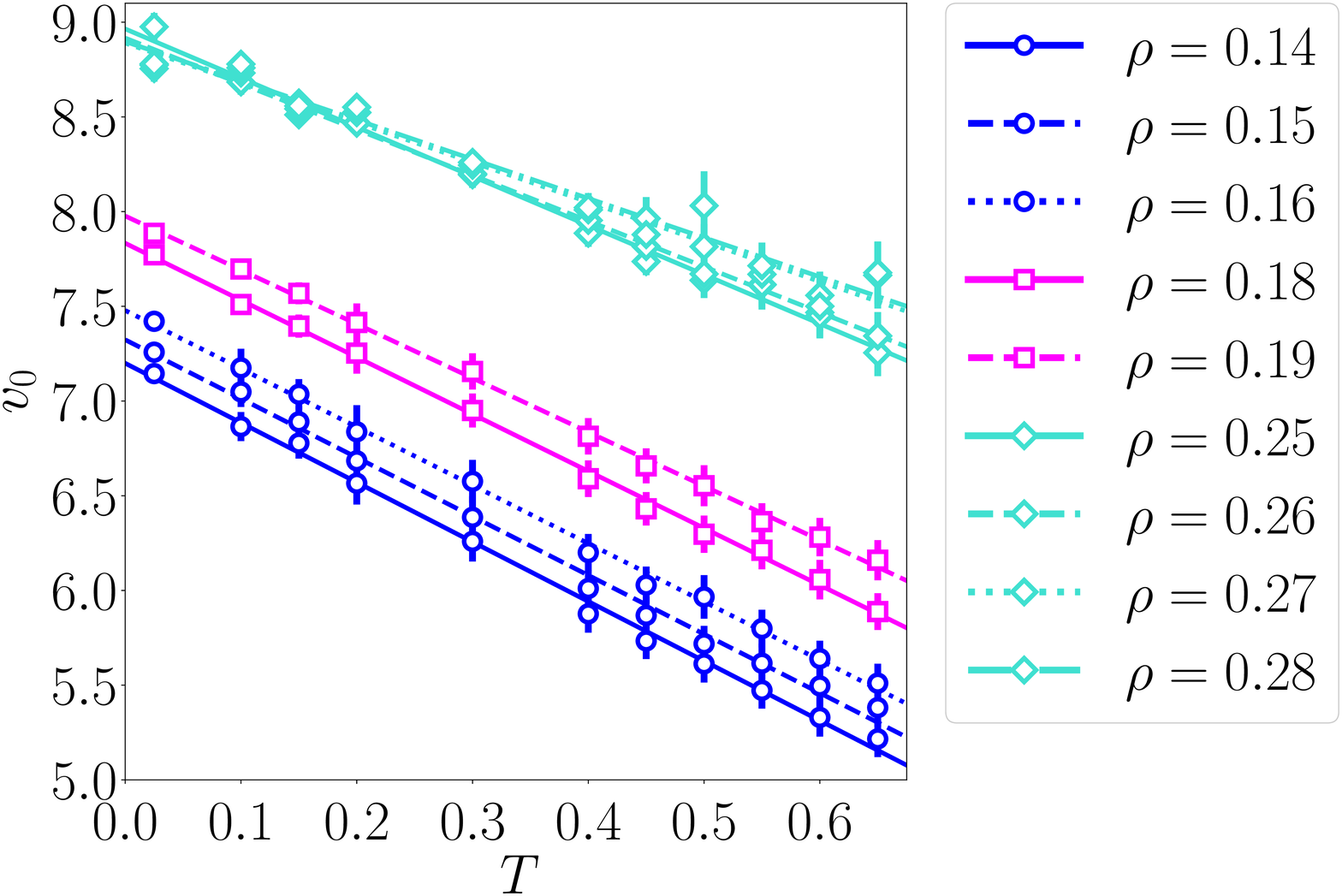}
  \end{center}
  \caption{(Color online) Logistic midpoint in function of temperature 
for several densities, where the 
           lines are linear fits. The blue circles are systems with five 
layers, magenta 
           squares are systems with four layers, and turquoise diamonds
 are systems with three 
           layers.}
  \label{ref_fig8}
\end{figure}

Next, we test if the behavior of $v_0$ with the temperature, for fixed 
density, is also 
affected by the number of layers. The figure~\ref{ref_fig3} illustrates
the behavior of the density versus temperature. 
Three different density regions are identified in this figure.
At the region I, $\rho=0.14$, $0.15$, 
and $0.16$, the fluid forms five layers.
At the region II, $\rho=0.18$, and $0.19$, the system is accommodated 
in four layers. At the region III, $\rho=0.25$, $0.26$, $0.27$, and $0.28$, 
three layers are formed. 
The figure~\ref{ref_fig8} shows the behavior of the 
logistic midpoint as a function of the temperature for densities in
 the regions I (circles), II (squares)
and III (diamonds) identified in the 
figure~\ref{ref_fig3}.  The transition velocity decreases linearly with 
temperature for all densities
analyzed.  
Densities  in the same
region (equal number of layers)  have the same
slope in the $v_0$ versus 
temperature graph. For the region I  in the 
figure~\ref{ref_fig3} the 
slope is $b_{I} =-3.09(2)$,
for the region II, $b_{II} =-2.92(6)$, and for the 
region III, $b_{III}=-2.4(2)$.

\begin{figure}[h!]
  \begin{center}
    \includegraphics[height=2.7in]{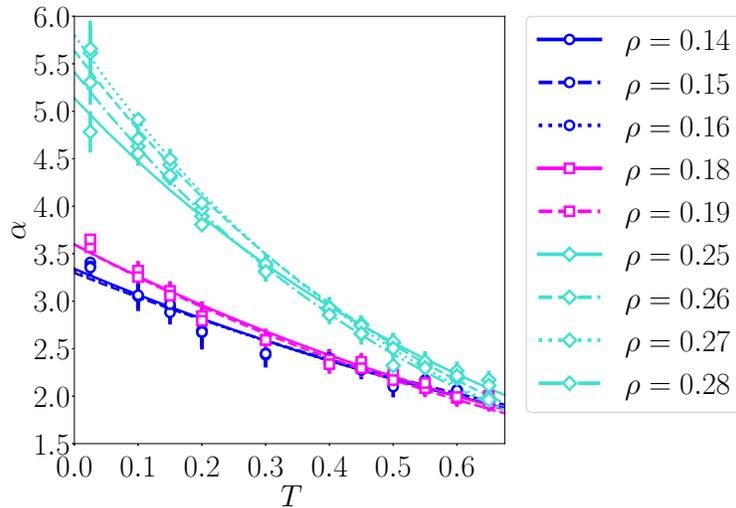}
  \end{center}
    \caption{(Color online) Logistic steepness in function of 
temperature for several densities, where the 
           curves are exponential fits. The blue circles are 
systems with five layers, magenta 
           squares are systems with four layers, and turquoise 
diamonds are systems with three 
           layers.}
  \label{ref_fig9}
\end{figure}

The figure~\ref{ref_fig9} shows the logistic 
steepness as a function
of the temperature for densities in regions I (circles), II 
(squares) and III (diamonds). 
It decreases exponentially with 
temperature, with an exponential fit 
($\alpha = \alpha_0\exp(-\lambda T)$). In the region I the mean 
exponential decay coefficient
is $\lambda_{I} =0.84(2)$, in the region II, is $\lambda_{II} =1.008(6)$, and 
in the region 
III, is $\lambda_{III}=1.62(7)$. Differently from  
the logistic midpoint, for high
temperatures all curves collapse. In this case the $\alpha$ value 
is almost the same for 
all densities. 

\begin{figure}[h!]
  \begin{center}
    \includegraphics[height=2.8in]{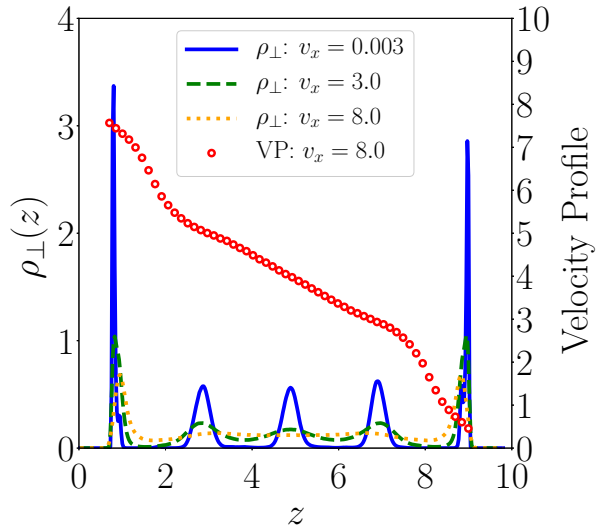}
  \end{center}
  \caption{(Color online) Left axis: transversal density profile for 
           $\rho_I=0.14$ at $T_I=0.025$ for one bottom wall 
           velocities in each boundary condition.
           Right axis: fluid velocity profile at the bottom wall 
           velocity $v_x=8.0$.}
  \label{ref_fig10}
\end{figure}

In order to understand how the transition to the 
global slip depends on the number of layers, 
the density profile of the system is analyzed
for different global densities and 
temperatures.
For  $\rho_I=0.14$ at $T_I=0.025$, as shown in the figure~\ref{ref_fig5}, 
the no-slip condition is valid for velocities below $2$, the
defect slip is valid for $2<v_x<v_0^I$, and
the global slip is valid for $v_x>v_0^I$, where $v_0^I=7.15(4)$.
The figure~\ref{ref_fig10} shows the dependence of the
transversal density profile 
with  $z$ for one bottom wall velocity in each boundary condition
for $\rho_I$ and $T_I$ (see figure~\ref{ref_fig3} for the location of
this point in the density versus temperature phase diagram).
The layering structure for different velocities in the no-slip condition
($v_x<2$) is very similar to the $v_x=0$ case, and exhibits 
layers without exchange of particles between them (solid line in 
figure~\ref{ref_fig10}).
For velocities in the defect slip condition the layers are present, but 
particles move between the layers 
(dashed line in figure~\ref{ref_fig10}).
For the global slip condition, at $v_x>v_0^I$,  the central layers 
are not present,  a uniform profile between 
the contact layers (dotted line in figure~\ref{ref_fig10}) is formed.
For all densities studied at $T_I=0.025$ we observe the same behavior 
seen in the figure~\ref{ref_fig10}. 
Figure~\ref{ref_fig10} also shows a fluid velocity profile
for  $\rho_I=0.14$ at $T_I=0.025$ when the plate velocity is $v_x=8.0$
 (circles).
Even when the system is in the global slip regime, the velocity profile 
is not
linear. This behavior is due to the structure in layers and indicates
the difficulty of an accurate determination for the slip length.

\begin{figure}[h!]
  \begin{center}
    \includegraphics[height=2.75in]{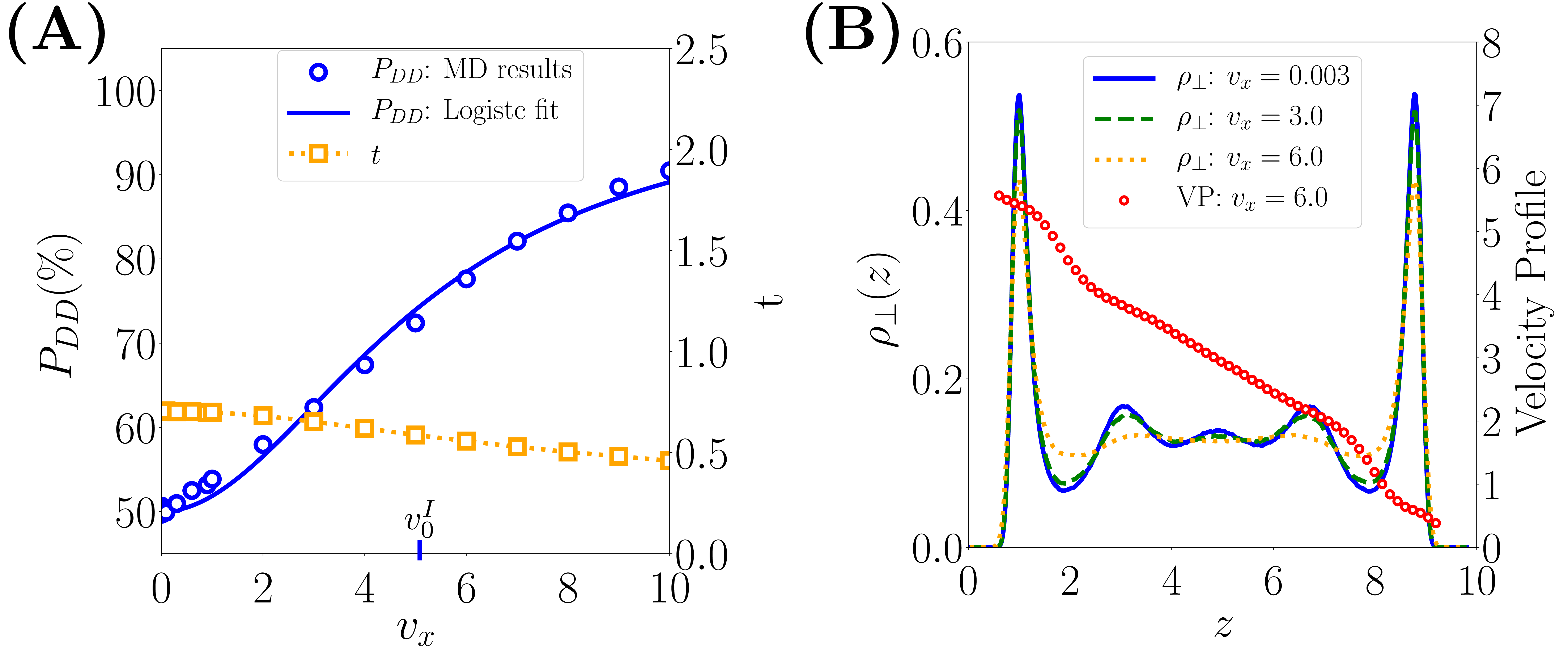}
  \end{center}
  \caption{(Color online) (A) Left axis: particles probability moving
            in the driven direction in function of bottom wall velocities. 
            The blue circles are results from simulations for $\rho_I = 0.14$
            at $T = 0.650$. The blue solid line is a logistic fit according eq. 
            \ref{ref_eq8} with $v_0^I = 5.2(1)$ and $\alpha^I=1.97(9)$. 
            Right axis: translational order parameter as a function of 
            bottom wall velocity for $\rho_I$ at $T=0.600$.
            (B) Left axis: transversal density profile for 
           $\rho_I=0.14$ at $T=0.650$ for one bottom wall 
           velocities in each boundary condition.
           Right axis: fluid velocity profile at the bottom wall 
           velocity $v_x=6.0$.}
  \label{ref_fig11}
\end{figure}

For the temperatures above the critical points in the 
figure~\ref{ref_fig3} the scenario is 
slightly different.
In this range 
of temperatures no transition 
is present even in the $v_x=0$ case, and the increase of $v_x$ 
promotes a smooth change in $P_{DD}$ as show in figure~\ref{ref_fig11} (A).
In this case the  increase in temperature 
promotes exchange of particles 
between the layers even for the no-slip case, as can be seen 
for $v_x=0.003$ (solid line) in figure~\ref{ref_fig11} (B).
Consequently, the defect slip appears for very small bottom wall 
velocities (figure~\ref{ref_fig11} (A)) with no 
significant change in the transversal density profile (dashed line, $v_x=3.0$, 
in figure~\ref{ref_fig11} (B)).
At the global slip the central layers are destroyed, 
and the system presents a bulk profile between the contact layers (dotted
line in figure~\ref{ref_fig11} (B)).
The velocity profile for high temperature continues to
show the nonlinear behavior observed at low temperature (see fig.~\ref{ref_fig11} (B)).

\begin{figure}[h!]
  \begin{center}
    \includegraphics[height=2.9in]{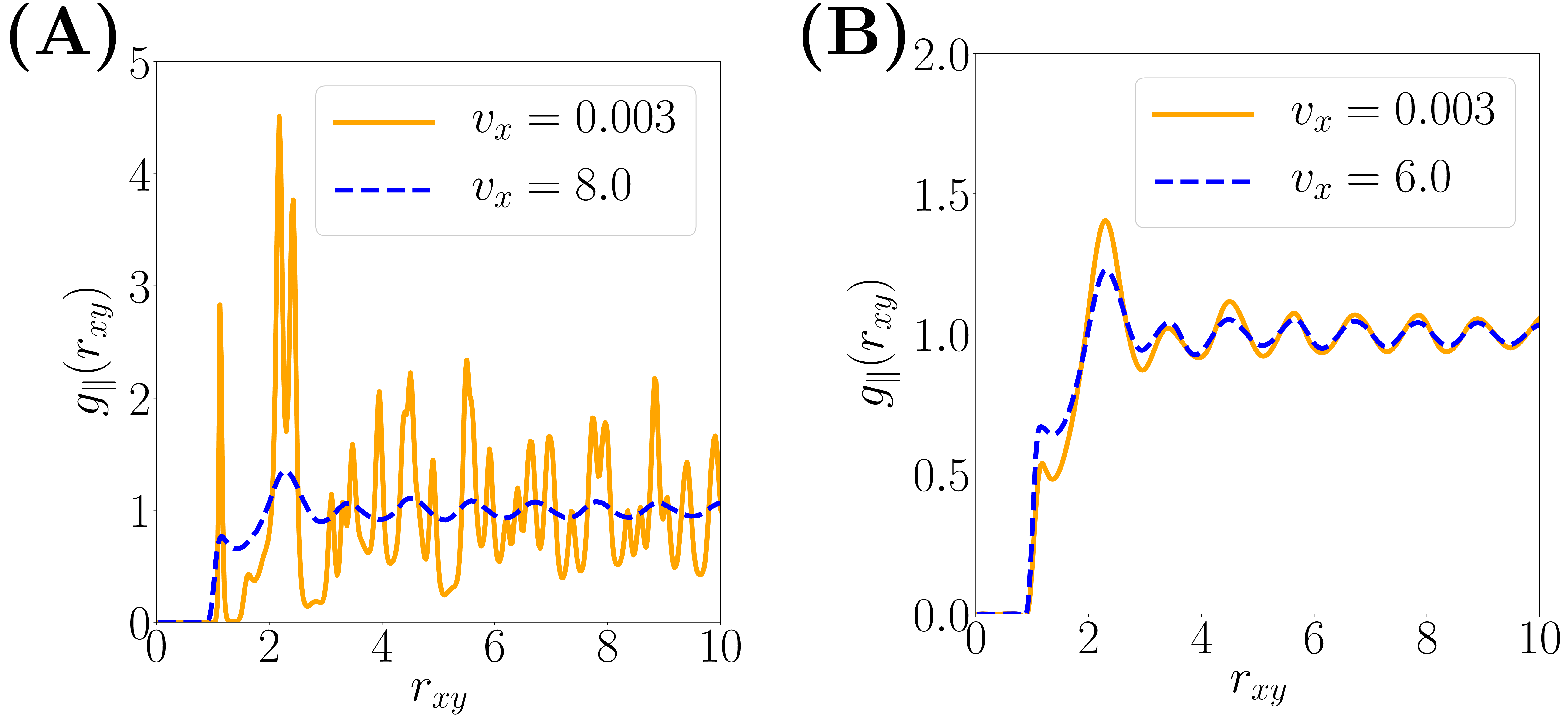}
  \end{center}
  \caption{(Color online) (A) Parallel radial distribution 
function of the contact layer for 
            $\rho_I=0.14$ at $T_I=0.025$ and (B) $T=0.650$.}
  \label{ref_fig12}
\end{figure}

Since the behavior of the
number of layers is affected by
$v_x$ quite differently 
when the system is the coexistence
region when compared with the 
supercritical region in the figure~\ref{ref_fig3},
the response of the structure to the 
change in the velocity rate is 
analyzed in detail for both regions.  
The figures~\ref{ref_fig5} and~\ref{ref_fig11} (A)
shows the translational order parameter (squares) 
as a function of 
bottom wall velocity for $\rho_I=0.14$. 
The translational order parameter, $t$, decreases
with increasing $v_x$, that is, the system 
becomes less structured with increasing shear 
level. In the coexistence region (fig.~\ref{ref_fig5}),
the decrease in $t$ value is much more pronounced than
in the supercritical region (fig.~\ref{ref_fig11} (A)).
The difference between the structures at low and high
shear level is evidenced in the parallel radial 
distribution function.
The figure~\ref{ref_fig12} 
shows the parallel radial 
distribution function 
of contact layer for $\rho_I=0.14$ at $T_I=0.025$ (coexistence region)(A) and $T=0.650$ (supercritical region)(B), for a
 low and a
high wall velocities. For low temperature case (coexistence region), the
 wall velocity leads to
 a transition
from an amorphous phase to a liquid phase at
 $v_x>v_0^I$ (figure~\ref{ref_fig12}
(A)). This behavior also was observed for $\rho_{II}=0.18$ and 
$\rho_{III} =0.25$ at
$T_I$ (coexistence region in the figure~\ref{ref_fig3}).  For the 
temperatures above the critical points,  the fluid is in 
the liquid
phase independent of the bottom wall 
velocities (figure~\ref{ref_fig12} (B)) and 
the $\alpha$ value 
is independent of the number of layers. This behavior 
also was observed
for $\rho_{II}$ and $\rho_{III}$ at $T=0.650$. 

\section{Conclusions}
In this work, we studied
the dynamical behavior of an water-like fluid under shear. 
As the wall speed
increases a transition from the defect to the global slip conditions
was observed. 

We showed that the defect slip appears due to an exchange of 
particles between the different fluid layers present
in the confined water-like fluid.

The dynamics of this exchange is defined by  the  bottom 
wall velocity and the temperature. For low temperatures,
 the velocity of the bottom wall required 
for the defect slip to occur is large
while  for high
 temperatures, this velocity is low.

The bottom wall velocity necessary to 
promote the global slip condition depends on  density,
temperature and number of fluid layers.
For a fixed density, the velocity of the bottom wall required
for the global slip to occur decreases linearly with increasing temperature,
with a  slope which depends on the number of fluid layers.

We also found that the transition between the no-slip to 
the global slip is more smooth 
for high temperatures where no phase transition is observed in contact layer.
In this situation the parameter $\alpha$ 
is independent of density.
For low temperatures the water-like fluid present different crystal-liquid 
phases, the parameter $\alpha$ present high values dependent of the number 
of fluid layers, and the the transition 
from no-slip to global slip is less smooth.

Our findings are consistent with the hypothesis that slip is dependent on 
temperature,
density and shear rate. However, our work shows  that although the slip 
is a dynamic phenomenon directly related to the contact layer, the behavior 
of the fluid between 
these layers is determinant for the occurrence or not of the slip.
Therefore, the anomalous dynamics of confined water-like fluid can be 
understood through the relation
of the occurrence of the slip at liquid-solid interface and the anomalous 
thermodynamic and structure
that water assumes under confinement.

\section{Acknowledgments}
We thank the Brazilian agencies CNPq and  INCT-FCx  for the
financial support.

\bibliography{BIB_article1}

\begin{thebibliography}{69}%
\makeatletter
\providecommand \@ifxundefined [1]{%
 \@ifx{#1\undefined}
}%
\providecommand \@ifnum [1]{%
 \ifnum #1\expandafter \@firstoftwo
 \else \expandafter \@secondoftwo
 \fi
}%
\providecommand \@ifx [1]{%
 \ifx #1\expandafter \@firstoftwo
 \else \expandafter \@secondoftwo
 \fi
}%
\providecommand \natexlab [1]{#1}%
\providecommand \enquote  [1]{``#1''}%
\providecommand \bibnamefont  [1]{#1}%
\providecommand \bibfnamefont [1]{#1}%
\providecommand \citenamefont [1]{#1}%
\providecommand \href@noop [0]{\@secondoftwo}%
\providecommand \href [0]{\begingroup \@sanitize@url \@href}%
\providecommand \@href[1]{\@@startlink{#1}\@@href}%
\providecommand \@@href[1]{\endgroup#1\@@endlink}%
\providecommand \@sanitize@url [0]{\catcode `\\12\catcode `\$12\catcode
  `\&12\catcode `\#12\catcode `\^12\catcode `\_12\catcode `\%12\relax}%
\providecommand \@@startlink[1]{}%
\providecommand \@@endlink[0]{}%
\providecommand \url  [0]{\begingroup\@sanitize@url \@url }%
\providecommand \@url [1]{\endgroup\@href {#1}{\urlprefix }}%
\providecommand \urlprefix  [0]{URL }%
\providecommand \Eprint [0]{\href }%
\providecommand \doibase [0]{http://dx.doi.org/}%
\providecommand \selectlanguage [0]{\@gobble}%
\providecommand \bibinfo  [0]{\@secondoftwo}%
\providecommand \bibfield  [0]{\@secondoftwo}%
\providecommand \translation [1]{[#1]}%
\providecommand \BibitemOpen [0]{}%
\providecommand \bibitemStop [0]{}%
\providecommand \bibitemNoStop [0]{.\EOS\space}%
\providecommand \EOS [0]{\spacefactor3000\relax}%
\providecommand \BibitemShut  [1]{\csname bibitem#1\endcsname}%
\let\auto@bib@innerbib\@empty
\bibitem [{\citenamefont {Watanabe}\ \emph {et~al.}(1999)\citenamefont
  {Watanabe}, \citenamefont {Udagawa},\ and\ \citenamefont {Udagawa}}]{WAT99}%
  \BibitemOpen
  \bibfield  {author} {\bibinfo {author} {\bibfnamefont {K.}~\bibnamefont
  {Watanabe}}, \bibinfo {author} {\bibfnamefont {Y.}~\bibnamefont {Udagawa}}, \
  and\ \bibinfo {author} {\bibfnamefont {H.}~\bibnamefont {Udagawa}},\
  }\href@noop {} {\bibfield  {journal} {\bibinfo  {journal} {Journal of Fluid
  Mechanics}\ }\textbf {\bibinfo {volume} {381}},\ \bibinfo {pages} {225}
  (\bibinfo {year} {1999})}\BibitemShut {NoStop}%
\bibitem [{\citenamefont {Baudry}\ \emph {et~al.}(2001)\citenamefont {Baudry},
  \citenamefont {Charlaix}, \citenamefont {Tonck},\ and\ \citenamefont
  {Mazuyer}}]{BAU01}%
  \BibitemOpen
  \bibfield  {author} {\bibinfo {author} {\bibfnamefont {J.}~\bibnamefont
  {Baudry}}, \bibinfo {author} {\bibfnamefont {E.}~\bibnamefont {Charlaix}},
  \bibinfo {author} {\bibfnamefont {A.}~\bibnamefont {Tonck}}, \ and\ \bibinfo
  {author} {\bibfnamefont {D.}~\bibnamefont {Mazuyer}},\ }\href@noop {}
  {\bibfield  {journal} {\bibinfo  {journal} {Langmuir}\ }\textbf {\bibinfo
  {volume} {17}},\ \bibinfo {pages} {5232} (\bibinfo {year}
  {2001})}\BibitemShut {NoStop}%
\bibitem [{\citenamefont {Zhu}\ and\ \citenamefont {Granick}(2001)}]{ZHU01}%
  \BibitemOpen
  \bibfield  {author} {\bibinfo {author} {\bibfnamefont {Y.}~\bibnamefont
  {Zhu}}\ and\ \bibinfo {author} {\bibfnamefont {S.}~\bibnamefont {Granick}},\
  }\href@noop {} {\bibfield  {journal} {\bibinfo  {journal} {Physical Review
  Letters}\ }\textbf {\bibinfo {volume} {87}},\ \bibinfo {pages} {096105}
  (\bibinfo {year} {2001})}\BibitemShut {NoStop}%
\bibitem [{\citenamefont {Zhu}\ and\ \citenamefont {Granick}(2002)}]{ZHU02}%
  \BibitemOpen
  \bibfield  {author} {\bibinfo {author} {\bibfnamefont {Y.}~\bibnamefont
  {Zhu}}\ and\ \bibinfo {author} {\bibfnamefont {S.}~\bibnamefont {Granick}},\
  }\href@noop {} {\bibfield  {journal} {\bibinfo  {journal} {Physical Review
  Letters}\ }\textbf {\bibinfo {volume} {88}},\ \bibinfo {pages} {106102}
  (\bibinfo {year} {2002})}\BibitemShut {NoStop}%
\bibitem [{\citenamefont {Cottin-Bizonne}\ \emph {et~al.}(2005)\citenamefont
  {Cottin-Bizonne}, \citenamefont {Cross}, \citenamefont {Steinberger},\ and\
  \citenamefont {Charlaix}}]{COT05}%
  \BibitemOpen
  \bibfield  {author} {\bibinfo {author} {\bibfnamefont {C.}~\bibnamefont
  {Cottin-Bizonne}}, \bibinfo {author} {\bibfnamefont {B.}~\bibnamefont
  {Cross}}, \bibinfo {author} {\bibfnamefont {A.}~\bibnamefont {Steinberger}},
  \ and\ \bibinfo {author} {\bibfnamefont {E.}~\bibnamefont {Charlaix}},\
  }\href@noop {} {\bibfield  {journal} {\bibinfo  {journal} {Physical Review
  Letters}\ }\textbf {\bibinfo {volume} {94}},\ \bibinfo {pages} {056102}
  (\bibinfo {year} {2005})}\BibitemShut {NoStop}%
\bibitem [{\citenamefont {Neto}\ \emph {et~al.}(2005)\citenamefont {Neto},
  \citenamefont {Evans}, \citenamefont {Bonaccurso}, \citenamefont {Butt},\
  and\ \citenamefont {Craig}}]{NET05}%
  \BibitemOpen
  \bibfield  {author} {\bibinfo {author} {\bibfnamefont {C.}~\bibnamefont
  {Neto}}, \bibinfo {author} {\bibfnamefont {D.~R.}\ \bibnamefont {Evans}},
  \bibinfo {author} {\bibfnamefont {E.}~\bibnamefont {Bonaccurso}}, \bibinfo
  {author} {\bibfnamefont {H.-J.}\ \bibnamefont {Butt}}, \ and\ \bibinfo
  {author} {\bibfnamefont {V.~S.~J.}\ \bibnamefont {Craig}},\ }\href@noop {}
  {\bibfield  {journal} {\bibinfo  {journal} {Reports on Progress in Physics}\
  }\textbf {\bibinfo {volume} {68}},\ \bibinfo {pages} {2859} (\bibinfo {year}
  {2005})}\BibitemShut {NoStop}%
\bibitem [{\citenamefont {Joseph}\ and\ \citenamefont
  {Tabeling}(2005)}]{JOS05}%
  \BibitemOpen
  \bibfield  {author} {\bibinfo {author} {\bibfnamefont {P.}~\bibnamefont
  {Joseph}}\ and\ \bibinfo {author} {\bibfnamefont {P.}~\bibnamefont
  {Tabeling}},\ }\href@noop {} {\bibfield  {journal} {\bibinfo  {journal}
  {Physical Review E}\ }\textbf {\bibinfo {volume} {71}},\ \bibinfo {pages}
  {035303} (\bibinfo {year} {2005})}\BibitemShut {NoStop}%
\bibitem [{\citenamefont {Li}\ \emph {et~al.}(2014)\citenamefont {Li},
  \citenamefont {Mo},\ and\ \citenamefont {Li}}]{LI14}%
  \BibitemOpen
  \bibfield  {author} {\bibinfo {author} {\bibfnamefont {L.}~\bibnamefont
  {Li}}, \bibinfo {author} {\bibfnamefont {J.}~\bibnamefont {Mo}}, \ and\
  \bibinfo {author} {\bibfnamefont {Z.}~\bibnamefont {Li}},\ }\href@noop {}
  {\bibfield  {journal} {\bibinfo  {journal} {Physical Review E}\ }\textbf
  {\bibinfo {volume} {90}},\ \bibinfo {pages} {033003} (\bibinfo {year}
  {2014})}\BibitemShut {NoStop}%
\bibitem [{\citenamefont {Bocquet}\ and\ \citenamefont {Barrat}(1994)}]{BOC94}%
  \BibitemOpen
  \bibfield  {author} {\bibinfo {author} {\bibfnamefont {L.}~\bibnamefont
  {Bocquet}}\ and\ \bibinfo {author} {\bibfnamefont {J.-L.}\ \bibnamefont
  {Barrat}},\ }\href@noop {} {\bibfield  {journal} {\bibinfo  {journal}
  {Physical Review E}\ }\textbf {\bibinfo {volume} {49}},\ \bibinfo {pages}
  {3079} (\bibinfo {year} {1994})}\BibitemShut {NoStop}%
\bibitem [{\citenamefont {Vinogradova}(1995)}]{VIN95}%
  \BibitemOpen
  \bibfield  {author} {\bibinfo {author} {\bibfnamefont {O.~I.}\ \bibnamefont
  {Vinogradova}},\ }\href@noop {} {\bibfield  {journal} {\bibinfo  {journal}
  {Langmuir}\ }\textbf {\bibinfo {volume} {11}},\ \bibinfo {pages} {2213}
  (\bibinfo {year} {1995})}\BibitemShut {NoStop}%
\bibitem [{\citenamefont {Sokhan}\ \emph {et~al.}(2001)\citenamefont {Sokhan},
  \citenamefont {Nicholson},\ and\ \citenamefont {Quirke}}]{SOK01}%
  \BibitemOpen
  \bibfield  {author} {\bibinfo {author} {\bibfnamefont {V.}~\bibnamefont
  {Sokhan}}, \bibinfo {author} {\bibfnamefont {D.}~\bibnamefont {Nicholson}}, \
  and\ \bibinfo {author} {\bibfnamefont {N.}~\bibnamefont {Quirke}},\
  }\href@noop {} {\bibfield  {journal} {\bibinfo  {journal} {The Journal of
  Chemical Physics}\ }\textbf {\bibinfo {volume} {115}},\ \bibinfo {pages}
  {3878} (\bibinfo {year} {2001})}\BibitemShut {NoStop}%
\bibitem [{\citenamefont {Bocquet}\ and\ \citenamefont {Barrat}(2007)}]{BOC07}%
  \BibitemOpen
  \bibfield  {author} {\bibinfo {author} {\bibfnamefont {L.}~\bibnamefont
  {Bocquet}}\ and\ \bibinfo {author} {\bibfnamefont {J.-L.}\ \bibnamefont
  {Barrat}},\ }\href@noop {} {\bibfield  {journal} {\bibinfo  {journal} {Soft
  Matter}\ }\textbf {\bibinfo {volume} {3}},\ \bibinfo {pages} {685} (\bibinfo
  {year} {2007})}\BibitemShut {NoStop}%
\bibitem [{\citenamefont {Niavarani}\ and\ \citenamefont
  {Priezjev}(2010)}]{NIA10}%
  \BibitemOpen
  \bibfield  {author} {\bibinfo {author} {\bibfnamefont {A.}~\bibnamefont
  {Niavarani}}\ and\ \bibinfo {author} {\bibfnamefont {N.~V.}\ \bibnamefont
  {Priezjev}},\ }\href@noop {} {\bibfield  {journal} {\bibinfo  {journal}
  {Physical Review E}\ }\textbf {\bibinfo {volume} {81}},\ \bibinfo {pages}
  {011606} (\bibinfo {year} {2010})}\BibitemShut {NoStop}%
\bibitem [{\citenamefont {Bhadauria}\ \emph {et~al.}(2015)\citenamefont
  {Bhadauria}, \citenamefont {Sanghi},\ and\ \citenamefont {Aluru}}]{BHA15}%
  \BibitemOpen
  \bibfield  {author} {\bibinfo {author} {\bibfnamefont {R.}~\bibnamefont
  {Bhadauria}}, \bibinfo {author} {\bibfnamefont {T.}~\bibnamefont {Sanghi}}, \
  and\ \bibinfo {author} {\bibfnamefont {N.~R.}\ \bibnamefont {Aluru}},\
  }\href@noop {} {\bibfield  {journal} {\bibinfo  {journal} {The Journal of
  Chemical Physics}\ }\textbf {\bibinfo {volume} {143}},\ \bibinfo {pages}
  {174702} (\bibinfo {year} {2015})}\BibitemShut {NoStop}%
\bibitem [{\citenamefont {Wagemann}\ \emph {et~al.}(2017)\citenamefont
  {Wagemann}, \citenamefont {Oyarzua}, \citenamefont {Walther},\ and\
  \citenamefont {Zambrano}}]{WAG17}%
  \BibitemOpen
  \bibfield  {author} {\bibinfo {author} {\bibfnamefont {E.}~\bibnamefont
  {Wagemann}}, \bibinfo {author} {\bibfnamefont {E.}~\bibnamefont {Oyarzua}},
  \bibinfo {author} {\bibfnamefont {J.~H.}\ \bibnamefont {Walther}}, \ and\
  \bibinfo {author} {\bibfnamefont {H.~A.}\ \bibnamefont {Zambrano}},\
  }\href@noop {} {\bibfield  {journal} {\bibinfo  {journal} {Physical Chemistry
  Chemical Physics}\ }\textbf {\bibinfo {volume} {19}},\ \bibinfo {pages}
  {8646} (\bibinfo {year} {2017})}\BibitemShut {NoStop}%
\bibitem [{\citenamefont {Lichter}\ \emph {et~al.}(2004)\citenamefont
  {Lichter}, \citenamefont {Roxin},\ and\ \citenamefont {Mandre}}]{LIC04}%
  \BibitemOpen
  \bibfield  {author} {\bibinfo {author} {\bibfnamefont {S.}~\bibnamefont
  {Lichter}}, \bibinfo {author} {\bibfnamefont {A.}~\bibnamefont {Roxin}}, \
  and\ \bibinfo {author} {\bibfnamefont {S.}~\bibnamefont {Mandre}},\
  }\href@noop {} {\bibfield  {journal} {\bibinfo  {journal} {Physical Review
  Letters}\ }\textbf {\bibinfo {volume} {93}},\ \bibinfo {pages} {086001}
  (\bibinfo {year} {2004})}\BibitemShut {NoStop}%
\bibitem [{\citenamefont {Lichter}\ \emph {et~al.}(2007)\citenamefont
  {Lichter}, \citenamefont {Martini}, \citenamefont {Snurr},\ and\
  \citenamefont {Wang}}]{LIC07}%
  \BibitemOpen
  \bibfield  {author} {\bibinfo {author} {\bibfnamefont {S.}~\bibnamefont
  {Lichter}}, \bibinfo {author} {\bibfnamefont {A.}~\bibnamefont {Martini}},
  \bibinfo {author} {\bibfnamefont {R.~Q.}\ \bibnamefont {Snurr}}, \ and\
  \bibinfo {author} {\bibfnamefont {Q.}~\bibnamefont {Wang}},\ }\href@noop {}
  {\bibfield  {journal} {\bibinfo  {journal} {Physical Review Letters}\
  }\textbf {\bibinfo {volume} {98}},\ \bibinfo {pages} {226001} (\bibinfo
  {year} {2007})}\BibitemShut {NoStop}%
\bibitem [{\citenamefont {Martini}\ \emph
  {et~al.}(2008{\natexlab{a}})\citenamefont {Martini}, \citenamefont {Hsu},
  \citenamefont {Patankar},\ and\ \citenamefont {Lichter}}]{MAR08I}%
  \BibitemOpen
  \bibfield  {author} {\bibinfo {author} {\bibfnamefont {A.}~\bibnamefont
  {Martini}}, \bibinfo {author} {\bibfnamefont {H.-Y.}\ \bibnamefont {Hsu}},
  \bibinfo {author} {\bibfnamefont {N.~A.}\ \bibnamefont {Patankar}}, \ and\
  \bibinfo {author} {\bibfnamefont {S.}~\bibnamefont {Lichter}},\ }\href@noop
  {} {\bibfield  {journal} {\bibinfo  {journal} {Physical Review Letters}\
  }\textbf {\bibinfo {volume} {100}},\ \bibinfo {pages} {206001} (\bibinfo
  {year} {2008}{\natexlab{a}})}\BibitemShut {NoStop}%
\bibitem [{\citenamefont {Martini}\ \emph
  {et~al.}(2008{\natexlab{b}})\citenamefont {Martini}, \citenamefont {Roxin},
  \citenamefont {Snurr}, \citenamefont {Wang},\ and\ \citenamefont
  {Lichter}}]{MAR08II}%
  \BibitemOpen
  \bibfield  {author} {\bibinfo {author} {\bibfnamefont {A.}~\bibnamefont
  {Martini}}, \bibinfo {author} {\bibfnamefont {A.}~\bibnamefont {Roxin}},
  \bibinfo {author} {\bibfnamefont {R.~Q.}\ \bibnamefont {Snurr}}, \bibinfo
  {author} {\bibfnamefont {Q.}~\bibnamefont {Wang}}, \ and\ \bibinfo {author}
  {\bibfnamefont {S.}~\bibnamefont {Lichter}},\ }\href@noop {} {\bibfield
  {journal} {\bibinfo  {journal} {Journal of Fluid Mechanics}\ }\textbf
  {\bibinfo {volume} {600}},\ \bibinfo {pages} {257} (\bibinfo {year}
  {2008}{\natexlab{b}})}\BibitemShut {NoStop}%
\bibitem [{\citenamefont {Cheng}\ and\ \citenamefont {Giordano}(2002)}]{CHE02}%
  \BibitemOpen
  \bibfield  {author} {\bibinfo {author} {\bibfnamefont {J.-T.}\ \bibnamefont
  {Cheng}}\ and\ \bibinfo {author} {\bibfnamefont {N.}~\bibnamefont
  {Giordano}},\ }\href@noop {} {\bibfield  {journal} {\bibinfo  {journal}
  {Physical Review E}\ }\textbf {\bibinfo {volume} {65}},\ \bibinfo {pages}
  {031206} (\bibinfo {year} {2002})}\BibitemShut {NoStop}%
\bibitem [{\citenamefont {Eslami}\ and\ \citenamefont
  {Muller-Plathe}(2010)}]{ESL10}%
  \BibitemOpen
  \bibfield  {author} {\bibinfo {author} {\bibfnamefont {H.}~\bibnamefont
  {Eslami}}\ and\ \bibinfo {author} {\bibfnamefont {F.}~\bibnamefont
  {Muller-Plathe}},\ }\href@noop {} {\bibfield  {journal} {\bibinfo  {journal}
  {Journal of Physical Chemistry B}\ }\textbf {\bibinfo {volume} {114}},\
  \bibinfo {pages} {387} (\bibinfo {year} {2010})}\BibitemShut {NoStop}%
\bibitem [{\citenamefont {Thomas}\ and\ \citenamefont
  {McGaughey}(2008)}]{THO08}%
  \BibitemOpen
  \bibfield  {author} {\bibinfo {author} {\bibfnamefont {J.~A.}\ \bibnamefont
  {Thomas}}\ and\ \bibinfo {author} {\bibfnamefont {A.~J.~H.}\ \bibnamefont
  {McGaughey}},\ }\href@noop {} {\bibfield  {journal} {\bibinfo  {journal}
  {Nano Letters}\ }\textbf {\bibinfo {volume} {8}},\ \bibinfo {pages} {2788}
  (\bibinfo {year} {2008})}\BibitemShut {NoStop}%
\bibitem [{\citenamefont {Martini}\ \emph {et~al.}(2006)\citenamefont
  {Martini}, \citenamefont {Liu}, \citenamefont {Snurr},\ and\ \citenamefont
  {Wang}}]{MAR06}%
  \BibitemOpen
  \bibfield  {author} {\bibinfo {author} {\bibfnamefont {A.}~\bibnamefont
  {Martini}}, \bibinfo {author} {\bibfnamefont {Y.}~\bibnamefont {Liu}},
  \bibinfo {author} {\bibfnamefont {R.~Q.}\ \bibnamefont {Snurr}}, \ and\
  \bibinfo {author} {\bibfnamefont {Q.~J.}\ \bibnamefont {Wang}},\ }\href@noop
  {} {\bibfield  {journal} {\bibinfo  {journal} {Tribology Letters}\ }\textbf
  {\bibinfo {volume} {21}},\ \bibinfo {pages} {217} (\bibinfo {year}
  {2006})}\BibitemShut {NoStop}%
\bibitem [{\citenamefont {Tretheway}\ and\ \citenamefont
  {Meinhart}(2002)}]{TRE02}%
  \BibitemOpen
  \bibfield  {author} {\bibinfo {author} {\bibfnamefont {D.~C.}\ \bibnamefont
  {Tretheway}}\ and\ \bibinfo {author} {\bibfnamefont {C.~D.}\ \bibnamefont
  {Meinhart}},\ }\href@noop {} {\bibfield  {journal} {\bibinfo  {journal}
  {Physics of Fluids}\ }\textbf {\bibinfo {volume} {14}},\ \bibinfo {pages}
  {L9} (\bibinfo {year} {2002})}\BibitemShut {NoStop}%
\bibitem [{\citenamefont {Bonaccurso}\ \emph {et~al.}(2002)\citenamefont
  {Bonaccurso}, \citenamefont {Kappl},\ and\ \citenamefont {Butt}}]{BON02}%
  \BibitemOpen
  \bibfield  {author} {\bibinfo {author} {\bibfnamefont {E.}~\bibnamefont
  {Bonaccurso}}, \bibinfo {author} {\bibfnamefont {M.}~\bibnamefont {Kappl}}, \
  and\ \bibinfo {author} {\bibfnamefont {H.-J.}\ \bibnamefont {Butt}},\
  }\href@noop {} {\bibfield  {journal} {\bibinfo  {journal} {Physical Review
  Letters}\ }\textbf {\bibinfo {volume} {88}},\ \bibinfo {pages} {076103}
  (\bibinfo {year} {2002})}\BibitemShut {NoStop}%
\bibitem [{\citenamefont {Choi}\ \emph {et~al.}(2003)\citenamefont {Choi},
  \citenamefont {Westin},\ and\ \citenamefont {Breuer}}]{CHO03}%
  \BibitemOpen
  \bibfield  {author} {\bibinfo {author} {\bibfnamefont {C.-H.}\ \bibnamefont
  {Choi}}, \bibinfo {author} {\bibfnamefont {K.~J.~A.}\ \bibnamefont {Westin}},
  \ and\ \bibinfo {author} {\bibfnamefont {K.~S.}\ \bibnamefont {Breuer}},\
  }\href@noop {} {\bibfield  {journal} {\bibinfo  {journal} {Physics of
  Fluids}\ }\textbf {\bibinfo {volume} {15}},\ \bibinfo {pages} {2897}
  (\bibinfo {year} {2003})}\BibitemShut {NoStop}%
\bibitem [{\citenamefont {Vinogradova}\ \emph {et~al.}(2009)\citenamefont
  {Vinogradova}, \citenamefont {Koynov}, \citenamefont {Best},\ and\
  \citenamefont {Feuillebois}}]{VIN09}%
  \BibitemOpen
  \bibfield  {author} {\bibinfo {author} {\bibfnamefont {O.~I.}\ \bibnamefont
  {Vinogradova}}, \bibinfo {author} {\bibfnamefont {K.}~\bibnamefont {Koynov}},
  \bibinfo {author} {\bibfnamefont {A.}~\bibnamefont {Best}}, \ and\ \bibinfo
  {author} {\bibfnamefont {F.}~\bibnamefont {Feuillebois}},\ }\href@noop {}
  {\bibfield  {journal} {\bibinfo  {journal} {Physical Review Letters}\
  }\textbf {\bibinfo {volume} {102}},\ \bibinfo {pages} {118302} (\bibinfo
  {year} {2009})}\BibitemShut {NoStop}%
\bibitem [{\citenamefont {Xue}\ \emph {et~al.}(2014)\citenamefont {Xue},
  \citenamefont {Wu}, \citenamefont {Pei}, \citenamefont {Duan}, \citenamefont
  {Xue},\ and\ \citenamefont {Zhou}}]{XUE14}%
  \BibitemOpen
  \bibfield  {author} {\bibinfo {author} {\bibfnamefont {Y.}~\bibnamefont
  {Xue}}, \bibinfo {author} {\bibfnamefont {Y.}~\bibnamefont {Wu}}, \bibinfo
  {author} {\bibfnamefont {X.}~\bibnamefont {Pei}}, \bibinfo {author}
  {\bibfnamefont {H.}~\bibnamefont {Duan}}, \bibinfo {author} {\bibfnamefont
  {Q.}~\bibnamefont {Xue}}, \ and\ \bibinfo {author} {\bibfnamefont
  {F.}~\bibnamefont {Zhou}},\ }\href@noop {} {\bibfield  {journal} {\bibinfo
  {journal} {Langmuir}\ }\textbf {\bibinfo {volume} {31}},\ \bibinfo {pages}
  {226} (\bibinfo {year} {2014})}\BibitemShut {NoStop}%
\bibitem [{\citenamefont {Majumder}\ \emph {et~al.}(2005)\citenamefont
  {Majumder}, \citenamefont {Chopra}, \citenamefont {Andrews},\ and\
  \citenamefont {Hinds}}]{MAJ05}%
  \BibitemOpen
  \bibfield  {author} {\bibinfo {author} {\bibfnamefont {M.}~\bibnamefont
  {Majumder}}, \bibinfo {author} {\bibfnamefont {N.}~\bibnamefont {Chopra}},
  \bibinfo {author} {\bibfnamefont {R.}~\bibnamefont {Andrews}}, \ and\
  \bibinfo {author} {\bibfnamefont {B.~J.}\ \bibnamefont {Hinds}},\ }\href@noop
  {} {\bibfield  {journal} {\bibinfo  {journal} {Nature}\ }\textbf {\bibinfo
  {volume} {438}},\ \bibinfo {pages} {44} (\bibinfo {year} {2005})}\BibitemShut
  {NoStop}%
\bibitem [{\citenamefont {Holt}\ \emph {et~al.}(2006)\citenamefont {Holt},
  \citenamefont {Park}, \citenamefont {Wang}, \citenamefont {Stadermann},
  \citenamefont {Artyukhin}, \citenamefont {Grigoropoulos}, \citenamefont
  {Noy},\ and\ \citenamefont {Bakajin}}]{HOL06}%
  \BibitemOpen
  \bibfield  {author} {\bibinfo {author} {\bibfnamefont {J.~K.}\ \bibnamefont
  {Holt}}, \bibinfo {author} {\bibfnamefont {H.~G.}\ \bibnamefont {Park}},
  \bibinfo {author} {\bibfnamefont {Y.}~\bibnamefont {Wang}}, \bibinfo {author}
  {\bibfnamefont {M.}~\bibnamefont {Stadermann}}, \bibinfo {author}
  {\bibfnamefont {A.~B.}\ \bibnamefont {Artyukhin}}, \bibinfo {author}
  {\bibfnamefont {C.~P.}\ \bibnamefont {Grigoropoulos}}, \bibinfo {author}
  {\bibfnamefont {A.}~\bibnamefont {Noy}}, \ and\ \bibinfo {author}
  {\bibfnamefont {O.}~\bibnamefont {Bakajin}},\ }\href@noop {} {\bibfield
  {journal} {\bibinfo  {journal} {Science}\ }\textbf {\bibinfo {volume}
  {312}},\ \bibinfo {pages} {1034} (\bibinfo {year} {2006})}\BibitemShut
  {NoStop}%
\bibitem [{\citenamefont {Qin}\ \emph {et~al.}(2011)\citenamefont {Qin},
  \citenamefont {Yuan}, \citenamefont {Zhao}, \citenamefont {Xie},\ and\
  \citenamefont {Liu}}]{QIN11}%
  \BibitemOpen
  \bibfield  {author} {\bibinfo {author} {\bibfnamefont {X.}~\bibnamefont
  {Qin}}, \bibinfo {author} {\bibfnamefont {Q.}~\bibnamefont {Yuan}}, \bibinfo
  {author} {\bibfnamefont {Y.}~\bibnamefont {Zhao}}, \bibinfo {author}
  {\bibfnamefont {S.}~\bibnamefont {Xie}}, \ and\ \bibinfo {author}
  {\bibfnamefont {Z.}~\bibnamefont {Liu}},\ }\href@noop {} {\bibfield
  {journal} {\bibinfo  {journal} {Nano Letters}\ }\textbf {\bibinfo {volume}
  {11}},\ \bibinfo {pages} {2173} (\bibinfo {year} {2011})}\BibitemShut
  {NoStop}%
\bibitem [{\citenamefont {Ternes}\ \emph {et~al.}(2017)\citenamefont {Ternes},
  \citenamefont {Mendoza-Coto},\ and\ \citenamefont {Salcedo}}]{TER17}%
  \BibitemOpen
  \bibfield  {author} {\bibinfo {author} {\bibfnamefont {P.}~\bibnamefont
  {Ternes}}, \bibinfo {author} {\bibfnamefont {A.}~\bibnamefont
  {Mendoza-Coto}}, \ and\ \bibinfo {author} {\bibfnamefont {E.}~\bibnamefont
  {Salcedo}},\ }\href@noop {} {\bibfield  {journal} {\bibinfo  {journal} {The
  Journal of Chemical Physics}\ }\textbf {\bibinfo {volume} {147}},\ \bibinfo
  {pages} {034510} (\bibinfo {year} {2017})}\BibitemShut {NoStop}%
\bibitem [{\citenamefont {Kannam}\ \emph {et~al.}(2012)\citenamefont {Kannam},
  \citenamefont {Todd}, \citenamefont {Hansen},\ and\ \citenamefont
  {Daivis}}]{KAN12}%
  \BibitemOpen
  \bibfield  {author} {\bibinfo {author} {\bibfnamefont {S.~K.}\ \bibnamefont
  {Kannam}}, \bibinfo {author} {\bibfnamefont {B.~D.}\ \bibnamefont {Todd}},
  \bibinfo {author} {\bibfnamefont {J.~S.}\ \bibnamefont {Hansen}}, \ and\
  \bibinfo {author} {\bibfnamefont {P.~J.}\ \bibnamefont {Daivis}},\
  }\href@noop {} {\bibfield  {journal} {\bibinfo  {journal} {The Journal of
  Chemical Physics}\ }\textbf {\bibinfo {volume} {136}},\ \bibinfo {pages}
  {024705} (\bibinfo {year} {2012})}\BibitemShut {NoStop}%
\bibitem [{\citenamefont {Koplik}\ \emph {et~al.}(1989)\citenamefont {Koplik},
  \citenamefont {Banavar},\ and\ \citenamefont {Willemsen}}]{KOP89}%
  \BibitemOpen
  \bibfield  {author} {\bibinfo {author} {\bibfnamefont {J.}~\bibnamefont
  {Koplik}}, \bibinfo {author} {\bibfnamefont {J.~R.}\ \bibnamefont {Banavar}},
  \ and\ \bibinfo {author} {\bibfnamefont {J.~F.}\ \bibnamefont {Willemsen}},\
  }\href@noop {} {\bibfield  {journal} {\bibinfo  {journal} {Physics of Fluids
  A: Fluid Dynamics}\ }\textbf {\bibinfo {volume} {1}},\ \bibinfo {pages} {781}
  (\bibinfo {year} {1989})}\BibitemShut {NoStop}%
\bibitem [{\citenamefont {Thompson}\ and\ \citenamefont
  {Robbins}(1990)}]{THO90}%
  \BibitemOpen
  \bibfield  {author} {\bibinfo {author} {\bibfnamefont {P.~A.}\ \bibnamefont
  {Thompson}}\ and\ \bibinfo {author} {\bibfnamefont {M.~O.}\ \bibnamefont
  {Robbins}},\ }\href@noop {} {\bibfield  {journal} {\bibinfo  {journal}
  {Physical Review A}\ }\textbf {\bibinfo {volume} {41}},\ \bibinfo {pages}
  {6830} (\bibinfo {year} {1990})}\BibitemShut {NoStop}%
\bibitem [{\citenamefont {Thompson}\ and\ \citenamefont
  {Troian}(1997)}]{THO97}%
  \BibitemOpen
  \bibfield  {author} {\bibinfo {author} {\bibfnamefont {P.~A.}\ \bibnamefont
  {Thompson}}\ and\ \bibinfo {author} {\bibfnamefont {S.~M.}\ \bibnamefont
  {Troian}},\ }\href@noop {} {\bibfield  {journal} {\bibinfo  {journal}
  {Nature}\ }\textbf {\bibinfo {volume} {389}},\ \bibinfo {pages} {360}
  (\bibinfo {year} {1997})}\BibitemShut {NoStop}%
\bibitem [{\citenamefont {Barrat}\ and\ \citenamefont {Bocquet}(1999)}]{BAR99}%
  \BibitemOpen
  \bibfield  {author} {\bibinfo {author} {\bibfnamefont {J.-L.}\ \bibnamefont
  {Barrat}}\ and\ \bibinfo {author} {\bibfnamefont {L.}~\bibnamefont
  {Bocquet}},\ }\href@noop {} {\bibfield  {journal} {\bibinfo  {journal}
  {Physical Review Letters}\ }\textbf {\bibinfo {volume} {82}},\ \bibinfo
  {pages} {4671} (\bibinfo {year} {1999})}\BibitemShut {NoStop}%
\bibitem [{\citenamefont {Pit}\ \emph {et~al.}(2000)\citenamefont {Pit},
  \citenamefont {Hervet},\ and\ \citenamefont {Leger}}]{PI00}%
  \BibitemOpen
  \bibfield  {author} {\bibinfo {author} {\bibfnamefont {R.}~\bibnamefont
  {Pit}}, \bibinfo {author} {\bibfnamefont {H.}~\bibnamefont {Hervet}}, \ and\
  \bibinfo {author} {\bibfnamefont {L.}~\bibnamefont {Leger}},\ }\href@noop {}
  {\bibfield  {journal} {\bibinfo  {journal} {Physical Review Letters}\
  }\textbf {\bibinfo {volume} {85}},\ \bibinfo {pages} {980} (\bibinfo {year}
  {2000})}\BibitemShut {NoStop}%
\bibitem [{\citenamefont {Cieplak}\ \emph {et~al.}(2001)\citenamefont
  {Cieplak}, \citenamefont {Koplik},\ and\ \citenamefont {Banavar}}]{CIE01}%
  \BibitemOpen
  \bibfield  {author} {\bibinfo {author} {\bibfnamefont {M.}~\bibnamefont
  {Cieplak}}, \bibinfo {author} {\bibfnamefont {J.}~\bibnamefont {Koplik}}, \
  and\ \bibinfo {author} {\bibfnamefont {J.~R.}\ \bibnamefont {Banavar}},\
  }\href@noop {} {\bibfield  {journal} {\bibinfo  {journal} {Physical Review
  Letters}\ }\textbf {\bibinfo {volume} {86}},\ \bibinfo {pages} {803}
  (\bibinfo {year} {2001})}\BibitemShut {NoStop}%
\bibitem [{\citenamefont {Joly}\ \emph {et~al.}(2006)\citenamefont {Joly},
  \citenamefont {Ybert},\ and\ \citenamefont {Bocquet}}]{JOL06}%
  \BibitemOpen
  \bibfield  {author} {\bibinfo {author} {\bibfnamefont {L.}~\bibnamefont
  {Joly}}, \bibinfo {author} {\bibfnamefont {C.}~\bibnamefont {Ybert}}, \ and\
  \bibinfo {author} {\bibfnamefont {L.}~\bibnamefont {Bocquet}},\ }\href@noop
  {} {\bibfield  {journal} {\bibinfo  {journal} {Physical Review Letters}\
  }\textbf {\bibinfo {volume} {96}},\ \bibinfo {pages} {046101} (\bibinfo
  {year} {2006})}\BibitemShut {NoStop}%
\bibitem [{\citenamefont {Priezjev}\ and\ \citenamefont
  {Troian}(2006)}]{PRI06}%
  \BibitemOpen
  \bibfield  {author} {\bibinfo {author} {\bibfnamefont {N.~V.}\ \bibnamefont
  {Priezjev}}\ and\ \bibinfo {author} {\bibfnamefont {S.~M.}\ \bibnamefont
  {Troian}},\ }\href@noop {} {\bibfield  {journal} {\bibinfo  {journal}
  {Journal of Fluid Mechanics}\ }\textbf {\bibinfo {volume} {554}},\ \bibinfo
  {pages} {25} (\bibinfo {year} {2006})}\BibitemShut {NoStop}%
\bibitem [{\citenamefont {Priezjev}(2007)}]{PRI07}%
  \BibitemOpen
  \bibfield  {author} {\bibinfo {author} {\bibfnamefont {N.~V.}\ \bibnamefont
  {Priezjev}},\ }\href@noop {} {\bibfield  {journal} {\bibinfo  {journal}
  {Physical Review E}\ }\textbf {\bibinfo {volume} {75}},\ \bibinfo {pages}
  {051605} (\bibinfo {year} {2007})}\BibitemShut {NoStop}%
\bibitem [{\citenamefont {Netz}\ \emph {et~al.}(2001)\citenamefont {Netz},
  \citenamefont {Starr}, \citenamefont {Stanley},\ and\ \citenamefont
  {Barbosa}}]{NET01}%
  \BibitemOpen
  \bibfield  {author} {\bibinfo {author} {\bibfnamefont {P.~A.}\ \bibnamefont
  {Netz}}, \bibinfo {author} {\bibfnamefont {F.~W.}\ \bibnamefont {Starr}},
  \bibinfo {author} {\bibfnamefont {H.~E.}\ \bibnamefont {Stanley}}, \ and\
  \bibinfo {author} {\bibfnamefont {M.~C.}\ \bibnamefont {Barbosa}},\
  }\href@noop {} {\bibfield  {journal} {\bibinfo  {journal} {The Journal of
  Chemical Physics}\ }\textbf {\bibinfo {volume} {115}},\ \bibinfo {pages}
  {344} (\bibinfo {year} {2001})}\BibitemShut {NoStop}%
\bibitem [{\citenamefont {Xu}\ \emph {et~al.}(2005)\citenamefont {Xu},
  \citenamefont {Kumar}, \citenamefont {Buldyrev}, \citenamefont {Chen},
  \citenamefont {Poole}, \citenamefont {Sciortino},\ and\ \citenamefont
  {Stanley}}]{XU05}%
  \BibitemOpen
  \bibfield  {author} {\bibinfo {author} {\bibfnamefont {L.}~\bibnamefont
  {Xu}}, \bibinfo {author} {\bibfnamefont {P.}~\bibnamefont {Kumar}}, \bibinfo
  {author} {\bibfnamefont {S.~V.}\ \bibnamefont {Buldyrev}}, \bibinfo {author}
  {\bibfnamefont {S.-H.}\ \bibnamefont {Chen}}, \bibinfo {author}
  {\bibfnamefont {P.~H.}\ \bibnamefont {Poole}}, \bibinfo {author}
  {\bibfnamefont {F.}~\bibnamefont {Sciortino}}, \ and\ \bibinfo {author}
  {\bibfnamefont {H.~E.}\ \bibnamefont {Stanley}},\ }\href@noop {} {\bibfield
  {journal} {\bibinfo  {journal} {Proceedings of the National Academy of
  Sciences of the United States of America}\ }\textbf {\bibinfo {volume}
  {102}},\ \bibinfo {pages} {16558} (\bibinfo {year} {2005})}\BibitemShut
  {NoStop}%
\bibitem [{\citenamefont {Xu}\ \emph {et~al.}(2006)\citenamefont {Xu},
  \citenamefont {Buldyrev}, \citenamefont {Angell},\ and\ \citenamefont
  {Stanley}}]{XU06}%
  \BibitemOpen
  \bibfield  {author} {\bibinfo {author} {\bibfnamefont {L.}~\bibnamefont
  {Xu}}, \bibinfo {author} {\bibfnamefont {S.~V.}\ \bibnamefont {Buldyrev}},
  \bibinfo {author} {\bibfnamefont {C.~A.}\ \bibnamefont {Angell}}, \ and\
  \bibinfo {author} {\bibfnamefont {H.~E.}\ \bibnamefont {Stanley}},\
  }\href@noop {} {\bibfield  {journal} {\bibinfo  {journal} {Physical Review
  E}\ }\textbf {\bibinfo {volume} {74}},\ \bibinfo {pages} {031108} (\bibinfo
  {year} {2006})}\BibitemShut {NoStop}%
\bibitem [{\citenamefont {de~Oliveira}\ \emph
  {et~al.}(2006{\natexlab{a}})\citenamefont {de~Oliveira}, \citenamefont
  {Netz}, \citenamefont {Colla},\ and\ \citenamefont {Barbosa}}]{OLI06I}%
  \BibitemOpen
  \bibfield  {author} {\bibinfo {author} {\bibfnamefont {A.~B.}\ \bibnamefont
  {de~Oliveira}}, \bibinfo {author} {\bibfnamefont {P.~A.}\ \bibnamefont
  {Netz}}, \bibinfo {author} {\bibfnamefont {T.}~\bibnamefont {Colla}}, \ and\
  \bibinfo {author} {\bibfnamefont {M.~C.}\ \bibnamefont {Barbosa}},\
  }\href@noop {} {\bibfield  {journal} {\bibinfo  {journal} {The Journal of
  Chemical Physics}\ }\textbf {\bibinfo {volume} {125}},\ \bibinfo {pages}
  {124503} (\bibinfo {year} {2006}{\natexlab{a}})}\BibitemShut {NoStop}%
\bibitem [{\citenamefont {de~Oliveira}\ \emph
  {et~al.}(2006{\natexlab{b}})\citenamefont {de~Oliveira}, \citenamefont
  {Netz}, \citenamefont {Colla},\ and\ \citenamefont {Barbosa}}]{OLI06II}%
  \BibitemOpen
  \bibfield  {author} {\bibinfo {author} {\bibfnamefont {A.~B.}\ \bibnamefont
  {de~Oliveira}}, \bibinfo {author} {\bibfnamefont {P.~A.}\ \bibnamefont
  {Netz}}, \bibinfo {author} {\bibfnamefont {T.}~\bibnamefont {Colla}}, \ and\
  \bibinfo {author} {\bibfnamefont {M.~C.}\ \bibnamefont {Barbosa}},\
  }\href@noop {} {\bibfield  {journal} {\bibinfo  {journal} {The Journal of
  Chemical Physics}\ }\textbf {\bibinfo {volume} {124}},\ \bibinfo {pages}
  {084505} (\bibinfo {year} {2006}{\natexlab{b}})}\BibitemShut {NoStop}%
\bibitem [{\citenamefont {Barraz}\ \emph {et~al.}(2009)\citenamefont {Barraz},
  \citenamefont {Salcedo},\ and\ \citenamefont {Barbosa}}]{BAR09}%
  \BibitemOpen
  \bibfield  {author} {\bibinfo {author} {\bibfnamefont {N.~M.}\ \bibnamefont
  {Barraz}, \bibfnamefont {Jr.}}, \bibinfo {author} {\bibfnamefont
  {E.}~\bibnamefont {Salcedo}}, \ and\ \bibinfo {author} {\bibfnamefont
  {M.~C.}\ \bibnamefont {Barbosa}},\ }\href@noop {} {\bibfield  {journal}
  {\bibinfo  {journal} {The Journal of Chemical Physics}\ }\textbf {\bibinfo
  {volume} {131}},\ \bibinfo {pages} {094504} (\bibinfo {year}
  {2009})}\BibitemShut {NoStop}%
\bibitem [{\citenamefont {Bordin}\ \emph
  {et~al.}(2014{\natexlab{a}})\citenamefont {Bordin}, \citenamefont {Krott},\
  and\ \citenamefont {Barbosa}}]{BOR14II}%
  \BibitemOpen
  \bibfield  {author} {\bibinfo {author} {\bibfnamefont {J.~R.}\ \bibnamefont
  {Bordin}}, \bibinfo {author} {\bibfnamefont {L.~B.}\ \bibnamefont {Krott}}, \
  and\ \bibinfo {author} {\bibfnamefont {M.~C.}\ \bibnamefont {Barbosa}},\
  }\href@noop {} {\bibfield  {journal} {\bibinfo  {journal} {The Journal of
  Chemical Physics}\ }\textbf {\bibinfo {volume} {141}},\ \bibinfo {pages}
  {144502} (\bibinfo {year} {2014}{\natexlab{a}})}\BibitemShut {NoStop}%
\bibitem [{\citenamefont {Bordin}\ \emph
  {et~al.}(2014{\natexlab{b}})\citenamefont {Bordin}, \citenamefont {Krott},\
  and\ \citenamefont {Barbosa}}]{BOR14I}%
  \BibitemOpen
  \bibfield  {author} {\bibinfo {author} {\bibfnamefont {J.~R.}\ \bibnamefont
  {Bordin}}, \bibinfo {author} {\bibfnamefont {L.~B.}\ \bibnamefont {Krott}}, \
  and\ \bibinfo {author} {\bibfnamefont {M.~C.}\ \bibnamefont {Barbosa}},\
  }\href@noop {} {\bibfield  {journal} {\bibinfo  {journal} {The Journal of
  Physical Chemistry C}\ }\textbf {\bibinfo {volume} {118}},\ \bibinfo {pages}
  {9497} (\bibinfo {year} {2014}{\natexlab{b}})}\BibitemShut {NoStop}%
\bibitem [{\citenamefont {Gavazzoni}\ \emph {et~al.}(2017)\citenamefont
  {Gavazzoni}, \citenamefont {Giovambattista}, \citenamefont {Netz},\ and\
  \citenamefont {Barbosa}}]{GAV17}%
  \BibitemOpen
  \bibfield  {author} {\bibinfo {author} {\bibfnamefont {C.}~\bibnamefont
  {Gavazzoni}}, \bibinfo {author} {\bibfnamefont {N.}~\bibnamefont
  {Giovambattista}}, \bibinfo {author} {\bibfnamefont {P.~A.}\ \bibnamefont
  {Netz}}, \ and\ \bibinfo {author} {\bibfnamefont {M.~C.}\ \bibnamefont
  {Barbosa}},\ }\href@noop {} {\bibfield  {journal} {\bibinfo  {journal} {The
  Journal of Chemical Physics}\ }\textbf {\bibinfo {volume} {146}},\ \bibinfo
  {pages} {234509} (\bibinfo {year} {2017})}\BibitemShut {NoStop}%
\bibitem [{\citenamefont {Neek-Amal}\ \emph {et~al.}(2016)\citenamefont
  {Neek-Amal}, \citenamefont {Peeters}, \citenamefont {Grigorieva},\ and\
  \citenamefont {Geim}}]{NEE16}%
  \BibitemOpen
  \bibfield  {author} {\bibinfo {author} {\bibfnamefont {M.}~\bibnamefont
  {Neek-Amal}}, \bibinfo {author} {\bibfnamefont {F.~M.}\ \bibnamefont
  {Peeters}}, \bibinfo {author} {\bibfnamefont {I.~V.}\ \bibnamefont
  {Grigorieva}}, \ and\ \bibinfo {author} {\bibfnamefont {A.~K.}\ \bibnamefont
  {Geim}},\ }\href {\doibase 10.1021/acsnano.6b00187} {\bibfield  {journal}
  {\bibinfo  {journal} {ACS Nano}\ }\textbf {\bibinfo {volume} {10}},\ \bibinfo
  {pages} {3685} (\bibinfo {year} {2016})}\BibitemShut {NoStop}%
\bibitem [{\citenamefont {Gallo}\ \emph {et~al.}(2010)\citenamefont {Gallo},
  \citenamefont {Rovere},\ and\ \citenamefont {Chen}}]{GAL10}%
  \BibitemOpen
  \bibfield  {author} {\bibinfo {author} {\bibfnamefont {P.}~\bibnamefont
  {Gallo}}, \bibinfo {author} {\bibfnamefont {M.}~\bibnamefont {Rovere}}, \
  and\ \bibinfo {author} {\bibfnamefont {S.-H.}\ \bibnamefont {Chen}},\
  }\href@noop {} {\bibfield  {journal} {\bibinfo  {journal} {Journal of
  Physics: Condensed Matter}\ }\textbf {\bibinfo {volume} {22}},\ \bibinfo
  {pages} {284102} (\bibinfo {year} {2010})}\BibitemShut {NoStop}%
\bibitem [{\citenamefont {Farimani}\ and\ \citenamefont {Aluru}(2016)}]{FAR16}%
  \BibitemOpen
  \bibfield  {author} {\bibinfo {author} {\bibfnamefont {A.~B.}\ \bibnamefont
  {Farimani}}\ and\ \bibinfo {author} {\bibfnamefont {N.~R.}\ \bibnamefont
  {Aluru}},\ }\href@noop {} {\bibfield  {journal} {\bibinfo  {journal} {The
  Journal of Physical Chemistry}\ }\textbf {\bibinfo {volume} {100}},\ \bibinfo
  {pages} {23763} (\bibinfo {year} {2016})}\BibitemShut {NoStop}%
\bibitem [{\citenamefont {De~Marzio}\ \emph {et~al.}(2017)\citenamefont
  {De~Marzio}, \citenamefont {Camisasca}, \citenamefont {Conde}, \citenamefont
  {Rovere},\ and\ \citenamefont {Gallo}}]{DEM17}%
  \BibitemOpen
  \bibfield  {author} {\bibinfo {author} {\bibfnamefont {M.}~\bibnamefont
  {De~Marzio}}, \bibinfo {author} {\bibfnamefont {G.}~\bibnamefont
  {Camisasca}}, \bibinfo {author} {\bibfnamefont {M.~M.}\ \bibnamefont
  {Conde}}, \bibinfo {author} {\bibfnamefont {M.}~\bibnamefont {Rovere}}, \
  and\ \bibinfo {author} {\bibfnamefont {P.}~\bibnamefont {Gallo}},\
  }\href@noop {} {\bibfield  {journal} {\bibinfo  {journal} {The Journal of
  Chemical Physics}\ }\textbf {\bibinfo {volume} {146}},\ \bibinfo {pages}
  {084505} (\bibinfo {year} {2017})}\BibitemShut {NoStop}%
\bibitem [{\citenamefont {Sega}\ \emph {et~al.}(2013)\citenamefont {Sega},
  \citenamefont {Sbragaglia}, \citenamefont {Biferale},\ and\ \citenamefont
  {Succi}}]{SEG13}%
  \BibitemOpen
  \bibfield  {author} {\bibinfo {author} {\bibfnamefont {M.}~\bibnamefont
  {Sega}}, \bibinfo {author} {\bibfnamefont {M.}~\bibnamefont {Sbragaglia}},
  \bibinfo {author} {\bibfnamefont {L.}~\bibnamefont {Biferale}}, \ and\
  \bibinfo {author} {\bibfnamefont {S.}~\bibnamefont {Succi}},\ }\href@noop {}
  {\bibfield  {journal} {\bibinfo  {journal} {Soft Matter}\ }\textbf {\bibinfo
  {volume} {9}},\ \bibinfo {pages} {8526} (\bibinfo {year} {2013})}\BibitemShut
  {NoStop}%
\bibitem [{\citenamefont {Jones}(1924)}]{JON24}%
  \BibitemOpen
  \bibfield  {author} {\bibinfo {author} {\bibfnamefont {J.~E.}\ \bibnamefont
  {Jones}},\ }\href@noop {} {\bibfield  {journal} {\bibinfo  {journal}
  {Proceedings of the Royal Society of London. Series A}\ }\textbf {\bibinfo
  {volume} {106}},\ \bibinfo {pages} {463} (\bibinfo {year}
  {1924})}\BibitemShut {NoStop}%
\bibitem [{\citenamefont {Lennard-Jones}(1931)}]{JON31}%
  \BibitemOpen
  \bibfield  {author} {\bibinfo {author} {\bibfnamefont {J.~E.}\ \bibnamefont
  {Lennard-Jones}},\ }\href@noop {} {\bibfield  {journal} {\bibinfo  {journal}
  {Proceedings of the Physical Society}\ }\textbf {\bibinfo {volume} {43}},\
  \bibinfo {pages} {461} (\bibinfo {year} {1931})}\BibitemShut {NoStop}%
\bibitem [{\citenamefont {Head-Gordon}\ and\ \citenamefont
  {Stillinger}(1993)}]{GOR93}%
  \BibitemOpen
  \bibfield  {author} {\bibinfo {author} {\bibfnamefont {T.}~\bibnamefont
  {Head-Gordon}}\ and\ \bibinfo {author} {\bibfnamefont {F.~H.}\ \bibnamefont
  {Stillinger}},\ }\href@noop {} {\bibfield  {journal} {\bibinfo  {journal}
  {The Journal of Chemical Physics}\ }\textbf {\bibinfo {volume} {98}},\
  \bibinfo {pages} {3313} (\bibinfo {year} {1993})}\BibitemShut {NoStop}%
\bibitem [{\citenamefont {Stillinger}\ and\ \citenamefont
  {Head-Gordon}(1993)}]{STI93}%
  \BibitemOpen
  \bibfield  {author} {\bibinfo {author} {\bibfnamefont {F.~H.}\ \bibnamefont
  {Stillinger}}\ and\ \bibinfo {author} {\bibfnamefont {T.}~\bibnamefont
  {Head-Gordon}},\ }\href@noop {} {\bibfield  {journal} {\bibinfo  {journal}
  {Physical Review E}\ }\textbf {\bibinfo {volume} {47}},\ \bibinfo {pages}
  {2484} (\bibinfo {year} {1993})}\BibitemShut {NoStop}%
\bibitem [{\citenamefont {Weeks}\ \emph {et~al.}(1971)\citenamefont {Weeks},
  \citenamefont {Chandler},\ and\ \citenamefont {Andersen}}]{WEE71}%
  \BibitemOpen
  \bibfield  {author} {\bibinfo {author} {\bibfnamefont {J.~D.}\ \bibnamefont
  {Weeks}}, \bibinfo {author} {\bibfnamefont {D.}~\bibnamefont {Chandler}}, \
  and\ \bibinfo {author} {\bibfnamefont {H.~C.}\ \bibnamefont {Andersen}},\
  }\href@noop {} {\bibfield  {journal} {\bibinfo  {journal} {The Journal of
  Chemical Physics}\ }\textbf {\bibinfo {volume} {54}},\ \bibinfo {pages}
  {5237} (\bibinfo {year} {1971})}\BibitemShut {NoStop}%
\bibitem [{\citenamefont {Frenkel}\ and\ \citenamefont {Smit}(1996)}]{FRE96}%
  \BibitemOpen
  \bibfield  {author} {\bibinfo {author} {\bibfnamefont {D.}~\bibnamefont
  {Frenkel}}\ and\ \bibinfo {author} {\bibfnamefont {B.}~\bibnamefont {Smit}},\
  }\href@noop {} {\emph {\bibinfo {title} {Understanding Molecular
  Simulation}}},\ \bibinfo {edition} {1st}\ ed.\ (\bibinfo  {publisher}
  {Academic, San Diego},\ \bibinfo {year} {1996})\BibitemShut {NoStop}%
\bibitem [{\citenamefont {Allen}\ and\ \citenamefont
  {Tildesley}(1987)}]{ALL87}%
  \BibitemOpen
  \bibfield  {author} {\bibinfo {author} {\bibfnamefont {M.~P.}\ \bibnamefont
  {Allen}}\ and\ \bibinfo {author} {\bibfnamefont {D.~J.}\ \bibnamefont
  {Tildesley}},\ }\href@noop {} {\emph {\bibinfo {title} {Computer Simulations
  of Liquids}}},\ \bibinfo {edition} {1st}\ ed.\ (\bibinfo  {publisher}
  {Claredon, Oxford},\ \bibinfo {year} {1987})\BibitemShut {NoStop}%
\bibitem [{\citenamefont {Kumar}\ \emph {et~al.}(2005)\citenamefont {Kumar},
  \citenamefont {Buldyrev}, \citenamefont {Starr}, \citenamefont
  {Giovambattista},\ and\ \citenamefont {Stanley}}]{KUM05}%
  \BibitemOpen
  \bibfield  {author} {\bibinfo {author} {\bibfnamefont {P.}~\bibnamefont
  {Kumar}}, \bibinfo {author} {\bibfnamefont {S.~V.}\ \bibnamefont {Buldyrev}},
  \bibinfo {author} {\bibfnamefont {F.~W.}\ \bibnamefont {Starr}}, \bibinfo
  {author} {\bibfnamefont {N.}~\bibnamefont {Giovambattista}}, \ and\ \bibinfo
  {author} {\bibfnamefont {H.~E.}\ \bibnamefont {Stanley}},\ }\href@noop {}
  {\bibfield  {journal} {\bibinfo  {journal} {Physical Review E}\ }\textbf
  {\bibinfo {volume} {72}},\ \bibinfo {pages} {051503} (\bibinfo {year}
  {2005})}\BibitemShut {NoStop}%
\bibitem [{\citenamefont {Kumar}\ \emph {et~al.}(2007)\citenamefont {Kumar},
  \citenamefont {Starr}, \citenamefont {Buldyrev},\ and\ \citenamefont
  {Stanley}}]{KUM07}%
  \BibitemOpen
  \bibfield  {author} {\bibinfo {author} {\bibfnamefont {P.}~\bibnamefont
  {Kumar}}, \bibinfo {author} {\bibfnamefont {F.~W.}\ \bibnamefont {Starr}},
  \bibinfo {author} {\bibfnamefont {S.~V.}\ \bibnamefont {Buldyrev}}, \ and\
  \bibinfo {author} {\bibfnamefont {H.~E.}\ \bibnamefont {Stanley}},\
  }\href@noop {} {\bibfield  {journal} {\bibinfo  {journal} {Physical Review
  E}\ }\textbf {\bibinfo {volume} {75}},\ \bibinfo {pages} {011202} (\bibinfo
  {year} {2007})}\BibitemShut {NoStop}%
\bibitem [{\citenamefont {Hoover}(1985)}]{HOO85}%
  \BibitemOpen
  \bibfield  {author} {\bibinfo {author} {\bibfnamefont {W.~G.}\ \bibnamefont
  {Hoover}},\ }\href@noop {} {\bibfield  {journal} {\bibinfo  {journal}
  {Physical Review A}\ }\textbf {\bibinfo {volume} {31}},\ \bibinfo {pages}
  {1695} (\bibinfo {year} {1985})}\BibitemShut {NoStop}%
\bibitem [{\citenamefont {Hoover}(1986)}]{HOO86}%
  \BibitemOpen
  \bibfield  {author} {\bibinfo {author} {\bibfnamefont {W.~G.}\ \bibnamefont
  {Hoover}},\ }\href@noop {} {\bibfield  {journal} {\bibinfo  {journal}
  {Physical Review A}\ }\textbf {\bibinfo {volume} {34}},\ \bibinfo {pages}
  {2499} (\bibinfo {year} {1986})}\BibitemShut {NoStop}%
\bibitem [{\citenamefont {Zangi}\ and\ \citenamefont {Rice}(2000)}]{ZAN00}%
  \BibitemOpen
  \bibfield  {author} {\bibinfo {author} {\bibfnamefont {R.}~\bibnamefont
  {Zangi}}\ and\ \bibinfo {author} {\bibfnamefont {S.~A.}\ \bibnamefont
  {Rice}},\ }\href@noop {} {\bibfield  {journal} {\bibinfo  {journal} {Physical
  Review E}\ }\textbf {\bibinfo {volume} {61}},\ \bibinfo {pages} {660}
  (\bibinfo {year} {2000})}\BibitemShut {NoStop}%
\bibitem [{\citenamefont {Errington}\ and\ \citenamefont
  {Debenedetti}(2001)}]{ERR01}%
  \BibitemOpen
  \bibfield  {author} {\bibinfo {author} {\bibfnamefont {J.~R.}\ \bibnamefont
  {Errington}}\ and\ \bibinfo {author} {\bibfnamefont {P.~G.}\ \bibnamefont
  {Debenedetti}},\ }\href@noop {} {\bibfield  {journal} {\bibinfo  {journal}
  {Nature}\ }\textbf {\bibinfo {volume} {409}},\ \bibinfo {pages} {318}
  (\bibinfo {year} {2001})}\BibitemShut {NoStop}%
\end{thebibliography}%
\end{document}